\documentclass[pdflatex,sn-mathphys-num]{sn-jnl}% Math and Physical Sciences Numbered Reference Style
\usepackage{epsfig}
\usepackage{psfrag,graphicx,color}
\usepackage{epstopdf}
\DeclareGraphicsExtensions{.pdf}
\AppendGraphicsExtensions{.tif}
\usepackage{graphics} % for pdf, bitmapped graphics files
\usepackage{epsfig} % for postscript graphics files

\usepackage{graphicx}%
\usepackage{multirow}%
\usepackage{amsmath,amssymb,amsfonts}%
\usepackage{amsthm}%
\usepackage{mathrsfs}%
\usepackage[title]{appendix}%
\usepackage{xcolor}%
\usepackage{textcomp}%
\usepackage{manyfoot}%
\usepackage{booktabs}%
\usepackage{algorithm}
\usepackage{algorithmic}
\usepackage{float}
\usepackage{listings}%
\usepackage{graphicx}
\usepackage{subcaption}
\usepackage{epsfig}

\theoremstyle{thmstyleone}%
\theoremstyle{thmstyletwo}%

\theoremstyle{thmstylethree}%

\begin{document}

\title[Article Title]{Dyna-Style Reinforcement Learning Modeling and Control of Non-linear Dynamics}

\author[1]{\fnm{Karim} \sur{Abdelsalam}}\email{karim.abdel-salam@guc.edu.eg}

\author[1]{\fnm{Zeyad} \sur{Gamal}}\email{zeyad.abdrabo@student.guc.edu.eg}
%\equalcont{These authors contributed equally to this work.}

\author[1]{\fnm{Ayman} \sur{El-Badawy}}\email{ayman.elbadawy@guc.edu.eg}
%\equalcont{These authors contributed equally to this work.}

\affil[1]{\orgdiv{Mechatronics Engineering Department}, \orgname{German University in Cairo}, \orgaddress{ \city{Cairo}, \country{Egypt}}}

%%=============================================================%%
%% GivenName	-> \fnm{Joergen W.}
%% Particle	-> \spfx{van der} -> surname prefix
%% FamilyName	-> \sur{Ploeg}
%% Suffix	-> \sfx{IV}
%% \author*[1,2]{\fnm{Joergen W.} \spfx{van der} \sur{Ploeg} 
%%  \sfx{IV}}\email{iauthor@gmail.com}
%%=============================================================%%

%%==================================%%
%% Sample for unstructured abstract %%
%%==================================%%

\abstract{Controlling systems with complex, nonlinear dynamics poses a significant challenge, particularly in achieving efficient and robust control. In this paper, we propose a Dyna-Style Reinforcement Learning control framework that integrates Sparse Identification of Nonlinear Dynamics (SINDy) with Twin Delayed Deep Deterministic Policy Gradient (TD3) reinforcement learning. SINDy is used to identify a data-driven model of the system, capturing its key dynamics without requiring an explicit physical model. This identified model is used to generate synthetic rollouts that are periodically injected into the reinforcement learning replay buffer during training on the real environment, enabling efficient policy learning with limited data available. By leveraging this hybrid approach, we mitigate the sample inefficiency of traditional model-free reinforcement learning methods while ensuring accurate control of nonlinear systems. To demonstrate the effectiveness of this framework, we apply it to a bi-rotor system as a case study, evaluating its performance in stabilization and trajectory tracking. The results show that our SINDy-TD3 approach achieves superior accuracy and robustness compared to direct reinforcement learning techniques, highlighting the potential of combining data-driven modeling with reinforcement learning for complex dynamical systems.}


%%\pacs[JEL Classification]{D8, H51}

%%\pacs[MSC Classification]{35A01, 65L10, 65L12, 65L20, 65L70}

\maketitle

\section{Introduction}\label{sec1}

Nonlinear systems are important in various engineering and scientific fields, characterized by complex dynamics and their dependencies between variables. Analysis and control of these systems are difficult because they frequently display behaviors like cross-coupling, nonlinearity, and multi-dimensional motion. Developing controllers for nonlinear systems to achieve stabilization and trajectory tracking, while maintaining stability under dynamic conditions, remains a critical area of research. Advanced modelling and control techniques are necessary to handle problems like nonlinearity, cross-coupling effects, and external disturbances due to the complexity of these systems.

Several studies have explored different traditional control strategies for nonlinear systems, such as PID control,  $H_{\infty}$ control \cite{Ragi}, sliding mode control\cite{Rashad2017}, and fuzzy logic controllers\cite{fuzzy}, have demonstrated effectiveness in various applications by explicitly relying on mathematical models of the system dynamics.

However, all the previous control techniques rely heavily on the mathematical model which might not be available or might be inaccurate in most cases. To address this limitation, model-free approaches, particularly those based on deep reinforcement learning (DRL), have emerged as powerful alternatives for controlling nonlinear systems. These methods bypass the need for an explicit system model by learning optimal control policies directly from data through interaction with the environment. DRL techniques such as Twin Delayed Deep Deterministic Policy Gradient (TD3) have been successfully applied to complex platforms like quadrotors and other multi-rotor systems \cite{shehabrl, Mokhtar2023, Zeyad}, achieving stabilization and trajectory tracking by learning mappings from full system states to control actions

Nevertheless, the previous model free techniques require a huge amount of data to train the agents in order to get a successful results in stabilization and trajectory tracking \cite{SINDYRL}. This poses a challenge in an environment where there is low data available for training or it is expensive to run the hardware for long times to get training data or to train the hardware directly using model free control techniques will be a challenge. One proposed solution for this is acquiring a data-driven model from the low-data regime available then use it as an environment for training DRL \cite{SINDYRL, sindyrl2022}.

In \cite{DataDrivenMPC}, a data driven model was proposed using the Gaussian Process and then incorporates it with MPC to control the trajectory tracking of a quad rotor. Also, \cite{SINDYquadrotor} made a data driven model using sparse identification of non-linear dynamics (SINDy) to model a quad rotor then an MPC controller was applied on the SINDy model to stabilize the quad rotor. The proposed SINDy-MPC was tested on several trajectory and it showed successful tracking for all of them. Also, \cite{hashem} has tested the SINDy algorithm on a magnetic levitation system resulting in an accurate model which captures its key dynamics.  Besides, \cite{SINDYMPC} performed a comparison between several techniques to acquire data driven models for different systems to test which one is more accurate. The techniques used was Neural Networks, Dynamic Mode Decomposition (DMD) and SINDy and the results showed that models acquired by SINDy were the best model to capture the system dynamics more accurate than any other technique with the smallest amount of data.

In this paper, we propose a control strategy based on a Dyna-style reinforcement learning architecture \cite{Sutton1998} that integrates Sparse Identification of Nonlinear Dynamics (SINDy) with Twin Delayed Deep Deterministic Policy Gradient (TD3) to efficiently control nonlinear systems. Unlike conventional deep reinforcement learning approaches that require extensive interaction with the environment, our method interleaves direct training of the TD3 agent on real system experience with periodic injection of synthetic experience generated by a data-driven SINDy model. The SINDy model is learned from real interaction data and serves as a lightweight surrogate for simulating additional dynamics. This hybrid approach improves sample efficiency and accelerates learning. The proposed method was validated on a bi-rotor laboratory setup—a nonlinear system with strong cross-coupling that mimics helicopter dynamics. The SINDy-TD3 framework is evaluated for both stabilization and trajectory tracking tasks, demonstrating accurate control performance while significantly reducing the number of required real interactions.

\section{Dyna-Style SINDY-TD3 Framework}

%\begin{figure}[H]
%    \centering
%    \includegraphics[width=0.75\linewidth]{Figures_eps/Fig1.eps}
%    \caption{Dyna-Style SINDy-TD3 Reinforcement Learning Architecture }
%    \label{fig:Dyna-style}
%\end{figure}

\begin{figure}[H]
    \centering
    \includegraphics[width=0.75\linewidth]{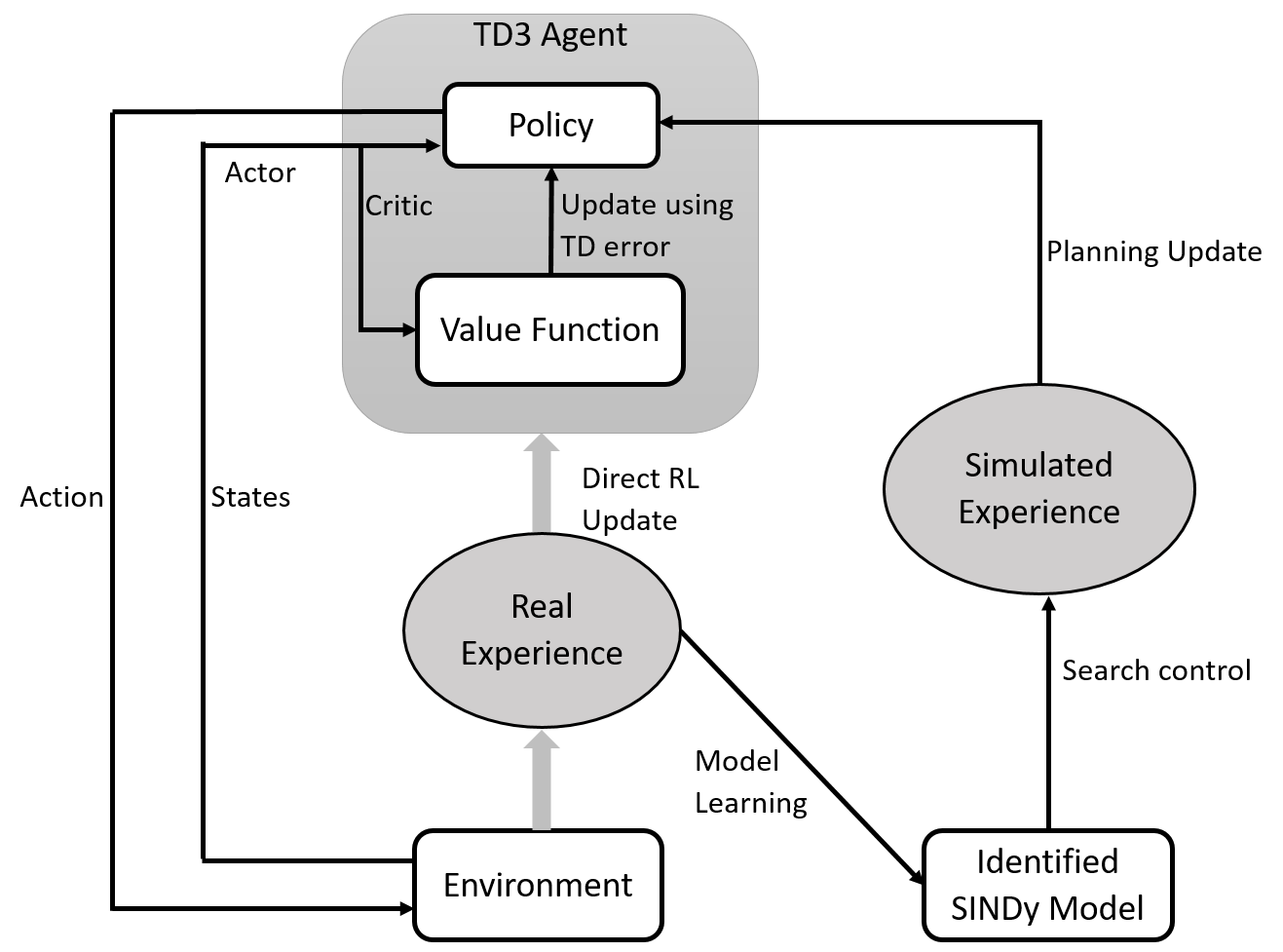}
    \caption{Dyna-Style SINDy-TD3 Reinforcement Learning Architecture }
    \label{fig:Dyna-style}
\end{figure}

%\begin{figure}[H]
%    \centering
%    \includegraphics[width=0.75\linewidth]{Figures_eps/Picture1.eps}
%    \caption{Dyna-Style SINDy-TD3 Reinforcement Learning Architecture }
%    \label{fig:Dyna-style}
%\end{figure}

This section introduces the proposed Dyna-Style SINDy-TD3 framework which aims to achieve data-driven control for non-linear systems with limited data available. The framework depends mainly on the integration of the main components which are Sparse Identification of Nonlinear Dynamics (SINDy) and Twin Delayed Deep Deterministic Policy Gradient (TD3).

This framework conceptually follows a Dyna-Q reinforcement learning architecture \cite{Sutton1998}. The Dyna-Q architecture is a foundational reinforcement learning framework that combines between model-based and model-free techniques to enhance data efficiency and maintain learning flexibility.  It depends on integrating three main components which are real-world interactions, model learning and planning. In Dyna-Q, an agent interacts with the environment to collect experience, which is used in two ways: directly to update the value function and policy using Q-Learning (direct reinforcement learning), and to build a model of the environment. This model is then used to generate simulated experience, which is treated as if it were real for further learning — a process known as planning.

The SINDy-TD3 framework proposed shown in Fig. \ref{fig:Dyna-style} follows a similar conceptual structure. In the beginning, real experience is collected from the environment using a default policy which can be a random input to the system, an untrained neural network or an input signal to explore and capture system dynamics \cite{SINDYRL}. Then, for the model learning phase, instead of using a simple table-based model as performed by the Dyna-Q, SINDy is used to learn a continuous data-driven model with limited interaction with real-world environment. This learned model will then act as a surrogate environment in order to generate simulated experience for the TD3 agent. In this framework, the learned dynamics model is used periodically to generate synthetic experience through simulated roll-outs using trained TD3 agent, these simulated experience is then added to the replay buffer along with real environment interactions done through direct RL so that the real buffer now has both real and simulated data. This combination of data will allow the TD3 agent to refine its policy between real-world episodes, improving sample efficiency and reducing the reliance on extensive physical interaction.

Algorithm \ref{alg:alg1} shows in detail the steps of the proposed framework. It starts with an initial data set provided by an untrained policy or a random input applied to the system. This data will then be used to train a SINDy model in order to capture the system dynamics. To ensure the learned model is accurate, the RL training reward will be computed using the SINDy model and the real environment and the error between them will be computed. If the error is high, the SINDy model will be updated with more TD3 real experience until the error is zero so that the model can be used in the planning step. While the model is learning, the TD3 agent will interact with the environment using the current policy to collect real experience and update the replay buffer in order to update the policy, meanwhile simulated experiences done by SINDy planning will also be added to the relay buffer in order to accelerate the training process until reaching the trained control policy which will then be deployed in the real environment.   

In the following sections, each step of the algorithm will be discussed in more details.

\begin{algorithm}[H]
\caption{Dyna-Style SINDy-TD3 Framework}\label{alg:dyna-sindy-td3}
\begin{algorithmic}[1]
\STATE \textbf{Input:} Initial dataset \( \mathcal{D} = \{(x_t, u_t, x_{t+1})\} \), RL hyperparameters \( \gamma, \tau, \sigma, N \), reward function \( r(x, u) \)
\STATE \textbf{Output:} Trained control policy \( \pi_\phi \)

\STATE \textbf{Data Collection:}  
Collect initial real-world trajectories \( \mathcal{D}_{\text{real}} \) from the physical system using an exploratory or random policy.

\STATE \textbf{Model Learning:}  
Train a SINDy model \( f_{\text{SINDy}}(x, u) \approx \dot{x} \) using \( \mathcal{D}_{\text{real}} \).

\STATE \textbf{Initialization:}  
Initialize replay buffer \( \mathcal{B} \leftarrow \mathcal{D}_{\text{real}} \), actor network \( \pi_\phi \), and critic networks \( Q_{\theta_1}, Q_{\theta_2} \).

\FOR{each episode}
    \STATE \textbf{Real Interaction:}  
    Interact with the real environment using the current policy \( \pi_\phi \) to collect transitions \( (x_t, u_t, r_t, x_{t+1}) \).

    \STATE Add collected real transitions to the replay buffer \( \mathcal{B} \).

    \STATE \textbf{Policy Learning:}  
    Sample mini-batches from \( \mathcal{B} \) and update the actor and critics using the TD3 algorithm.

    \STATE \textbf{Planning:}  
    \textbf{After every \( N \) episodes}, update the SINDy model using recent real transitions from \( \mathcal{B} \), simulate synthetic rollouts using the updated model to generate model-based transitions \( \mathcal{D}_{\text{model}} \), and add them to the replay buffer:
    \[
    \mathcal{B} \leftarrow \mathcal{B} \cup \mathcal{D}_{\text{model}}
    \]
\ENDFOR

\STATE \textbf{Deployment:}  
Deploy the final trained policy \( \pi_\phi \) on the real system for control and tracking.

\end{algorithmic}
\label{alg:alg1}
\end{algorithm}

\subsection{Training Data Collection}

For learning reliable and accurate models, particularly in nonlinear systems, high-quality data is essential. In order to ensure that the data collection process is thorough and not just a partial view, it must fully capture the dynamics of the system\cite{hashem}. 

Synthetic data can be generated to mimic real sensor measurements, enabling safe and controlled exploration of the system's response to various input profiles. Common input signals used for this purpose include chirp signals, which sweep through a range of frequencies to excite different dynamic modes, and Spatio-Periodic Harmonic Signals (SPHS), which are designed to capture a wide spectrum of behaviors with structured periodic content. The proper selection of such input signals is key to ensuring that the resulting dataset is informative enough for accurate model identification \cite{Book}.

\subsection{Model Learning}
 
After collecting the training data as shown in the previous section, it is formed into data matrix \cite{Book, SINDYMPC} as shown in equation \ref{eq:data}

\begin{equation} 
X = 
    \begin{bmatrix}
        x(t_1) & x(t_2) & x(t_3) & .... & x(t_n) 
    \end{bmatrix}^\top  
    \label{eq:data}
\end{equation}

and a similar matrix for the derivatives of the collected data

\begin{equation} 
\Dot{X} = 
    \begin{bmatrix}
        \Dot{x}(t_1) & \Dot{x}(t_2) & \Dot{x}(t_3) & .... & \Dot{x}(t_n) 
    \end{bmatrix}^\top  
\end{equation}

After that, a library of candidate non-linear functions should be constructed from the data in $X$ as shown in equation \ref{eq:library}:

\begin{equation}
\Phi(X) = 
    \begin{bmatrix}
        1 & X & X^2 & .... & X^n & .... & \sin{X} & \cos{X} & .... 
    \end{bmatrix}  
\label{eq:library}
\end{equation}

Where the matrix $\Phi(X)$ is a matrix with column vectors given by all possible time series of nth-degree polynomial in the state $x$ in addition to any function in terms of $X$. Generally, this library can be constructed using any functions that might be required for each dynamical system and can vary from one system to another. Now, the dynamics of the system can be represented in terms of the data matrices as shown in the following equation.

\begin{equation}
    \Dot{X} = \Phi(X) \Xi
\end{equation}

where each column $\xi_k$ in $\Xi$ is a vector of coefficients that determine the active terms that capture the dynamics of the system. Now, to provide an accurate model to really fit the system dynamics with the fewest possible terms in $\Xi$ we will need to identify it using a convex l1-regularized sparse regression:

\begin{equation}
    \xi_k = argmin ||\Dot{X}_k - \Theta(X)\xi_{k}^{'}||2 + \lambda ||\xi_{k}^{'}||
\label{eq:SINDY-EQ}
\end{equation}

As shown in equation \ref{eq:SINDY-EQ}, a key hyperparameter of the SINDy algorithm is the sparsity coefficient $\lambda$. This hyperparameter controls how sparse the identified model is. A sparser model with fewer terms in its dynamics is produced by a higher value of $\lambda$. This helps in producing a simplified model and reduces the chances of overfitting when there is noisy or limited data. However, an over-simplified model can result in emitting significant dynamics from the system. On the other hand, lower values of $\lambda$ can provide more detail in the produced model however, it can lead to over complicating the model. Thus, choosing an optimal value of $\lambda$ is crucial to find the balance between model accuracy and complexity which is in most cases should be a value between 0 and 1.

\subsection{Reinforcement Learning Control (TD3)}

The proposed control method which is used is Twin Delayed DDPG (TD3) algorithm \cite{TD3}. TD3 is an actor-critic model-free off policy algorithm. Actor-critic methods involve calculating a parameterized policy that functions as an actor. The actor's primary role is to determine the appropriate action based on the current state of the environment. Simultaneously, the method computes a value function that acts as a critic, assessing the actions selected by the actor. The critic also calculates the temporal difference (TD) error, which is used to update both the actor and the critic during the learning process. In TD3, it uses two critics instead of just one and take the lower value computed by both in order to reduce the overestimation bias commonly found in value based methods. Both the actor and critics are typically represented using Artificial Neural Networks (ANNs).

\begin{figure}[H]
    \centering
    \includegraphics[width=0.7\linewidth]{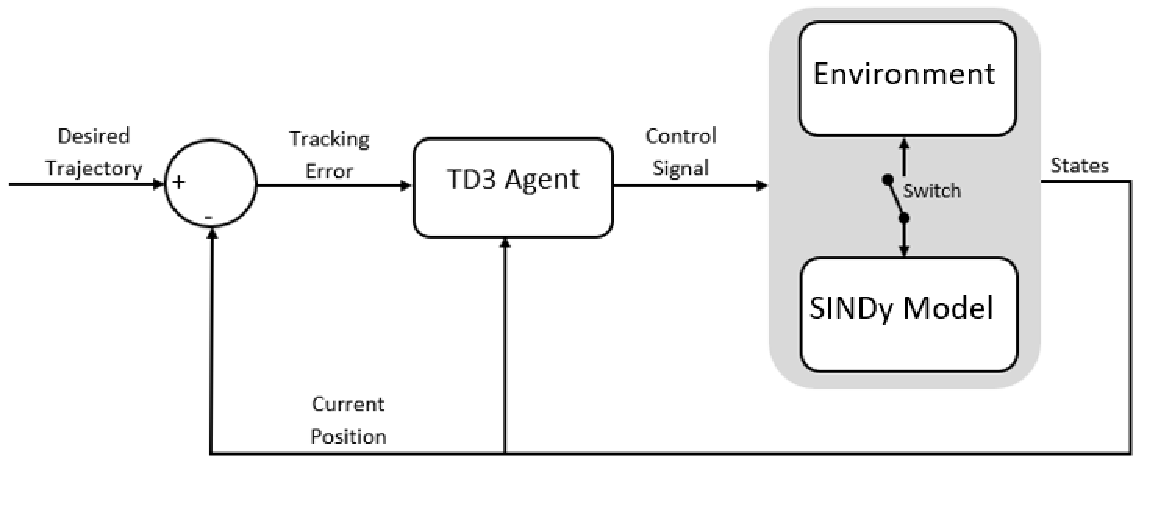}
    \caption{SINDy-TD3 control framework}
    \label{fig:control}
\end{figure}

The reinforcement learning control presented in this paper is shown in Fig. \ref{fig:control}. The framework is composed of several interconnected modules. The environment block represents the physical system under control and is responsible for generating data that reflects its true dynamics. A model learning block acquires this data to identify a simplified yet representative model of the system’s behavior, which can serve as a surrogate model for control policy training. This surrogate model is used to generate additional synthetic experiences added to the TD3 buffer to accelerate the learning process. Finally, the reinforcement learning agent block is trained using both real and synthetic data. The agent processes observations, including errors between desired and actual system outputs as well as other relevant state information, and produces control actions to drive the system toward its target behavior. This architecture allows for sample-efficient learning and improved performance in complex, nonlinear settings.

\section{Case Study: Bi-Rotor}

\begin{figure}[H]
    \centering
    \includegraphics[width=0.5\linewidth]{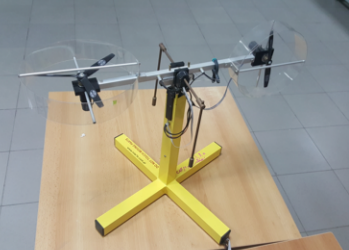}
    \caption{Bi-rotor lab setup \cite{Ragi}}
    \label{fig:lab-setup}
\end{figure}

In this section, the bi-rotor system  which is used as the case study to validate the proposed SINDy-TD3 technique will be explained in detail. The bi-rotor system consists of a beam with a rotor attached to each end. These rotors represent the main rotor,
which controls the pitch angle, and the tail rotor, which controls the azimuth angle. Each propeller is attached to a
DC-motor as is shown in Fig. \ref{fig:lab-setup}. Moreover, a counterbalance arm, with a mass attached to its end, is attached to the center of the beam.

\begin{figure}[H]
    \centering
    \includegraphics[width=0.7\linewidth]{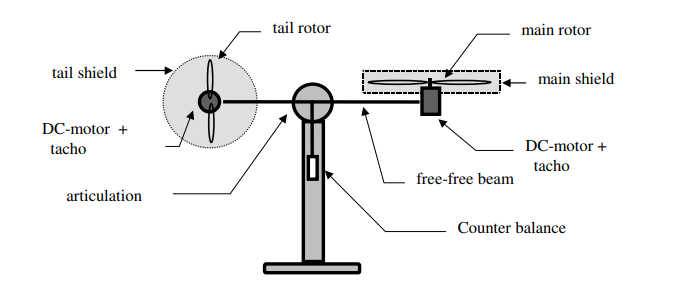}
    \caption{Bi-rotor schematic diagram \cite{Inteco2006}}
    \label{fig:TRMS}
\end{figure}

Fig. \ref{fig:TRMS} shows the side view of the bi-rotor. This system performs rotations in the azimuth and pitch directions. Subscripts $a$ and $p$ denote the azimuth and pitch directions respectively. The system consists of six states shown equation \ref{eq:states}.

\begin{equation} 
x = 
     \begin{bmatrix}
        x_1 \\ x_2 \\ x_3 \\ x_4 \\ x_5 \\ x_6
    \end{bmatrix}  
  = 
    \begin{bmatrix}
        \alpha_a \\ \alpha_p \\ \Omega_a  \\ \Omega_p \\  \omega_a \\ \omega_p 
    \end{bmatrix}
\label{eq:states}
\end{equation}

The state vector \( x \) shown in equation \ref{eq:states} of the bi-rotor system comprises six states, each representing specific properties of the vertical and horizontal rotors. The states \(  \alpha_a \) and \( \alpha_p \) correspond to the azimuth and pitch angles of the bi rotor respectively. The states \( \Omega_a \) and \( \Omega_p \) represent the azimuth and pitch angular velocities respectively. Finally, \(  \omega_a \) and \(  \omega_p \) represent the angular speed of the horizontal and vertical rotors, respectively \cite{Inteco2006}. These states were chosen as they are physically meaningful and should be controlled for effective stabilization of the bi-rotor.

\subsection{Data-Driven Model}

\subsubsection{Training Data Generation}

\begin{figure}[H]
    \centering
    \includegraphics[width=0.5\linewidth]{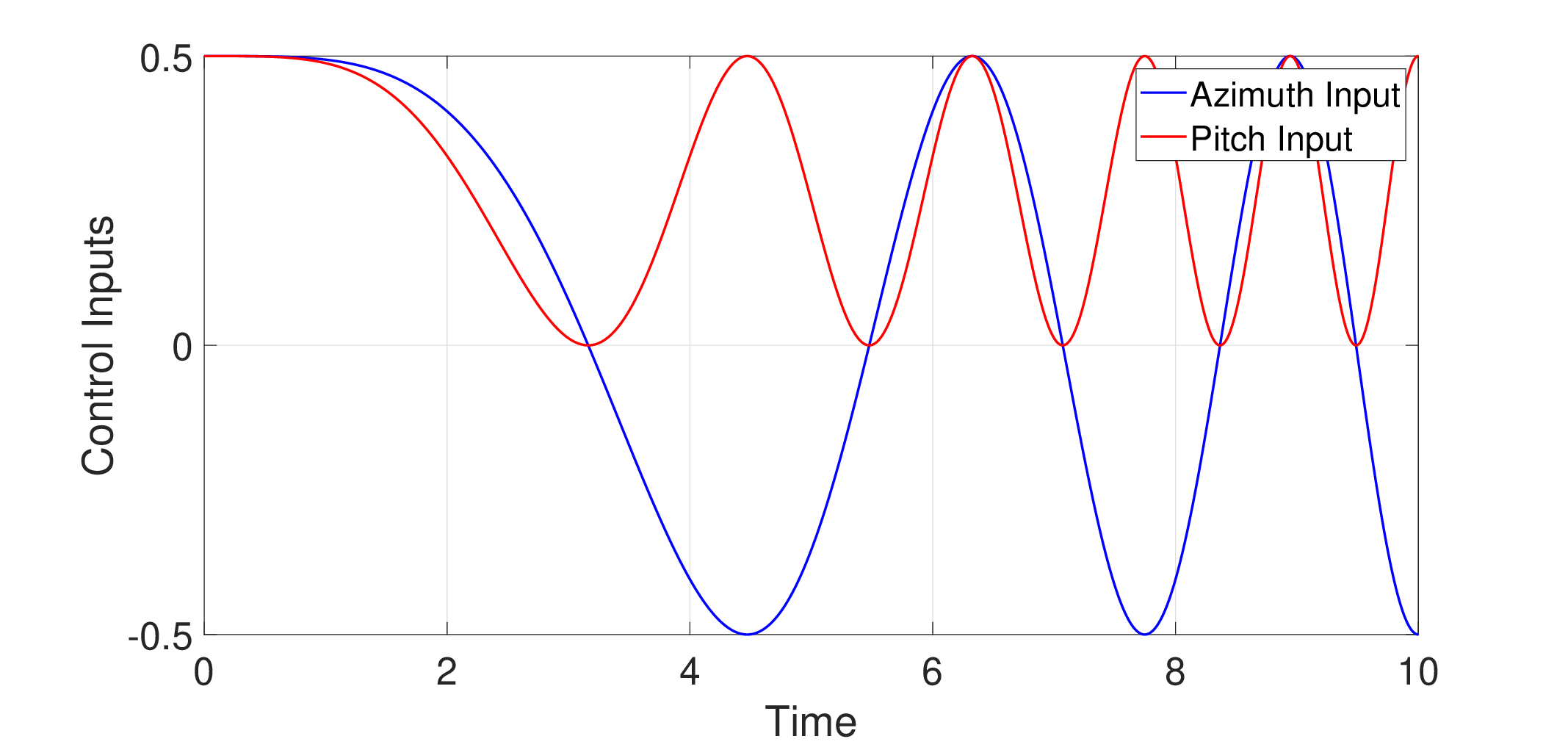}
    \caption{Chirp signal inputs to the Bi-rotor}
    \label{fig:chirp}
\end{figure}

The chirp signal shown in equation \ref{eq:chirp} was utilized as the actuation input to excite the bi-rotor system over a broad frequency range. Because it progressively sweeps through various frequencies, this input is perfect for system identification because it enables a thorough investigation of the dynamics of the system. The chirp signal in Fig. \ref{fig:chirp} provides a variable input, ensuring that the system's slow and fast dynamics are recorded. This allows for the collection of data that accurately depicts the bi-rotor's complete nonlinear behavior. Moreover, the chosen frequency range was carefully selected to ensure the system remained within the hardware's operational constraints, avoiding saturation or unsafe conditions during excitation.

% \begin{equation}
% u(t) = 0.5 \cdot \left( \text{chirp}(t, [], \max(t_{\text{span}}), 1) \right)^2
% \label{eq:chirp_signal}
% \end{equation}

\begin{equation}
u(t) = \text{chirp}(t) = \cos\left( \pi \cdot t \cdot \frac{1}{t_{\text{span}}} \cdot \left( 1 - \frac{t}{t_{\text{span}}} \right) \right)
\label{eq:chirp}
\end{equation}

\begin{figure}[H]
    \centering
    \includegraphics[width=0.5\linewidth]{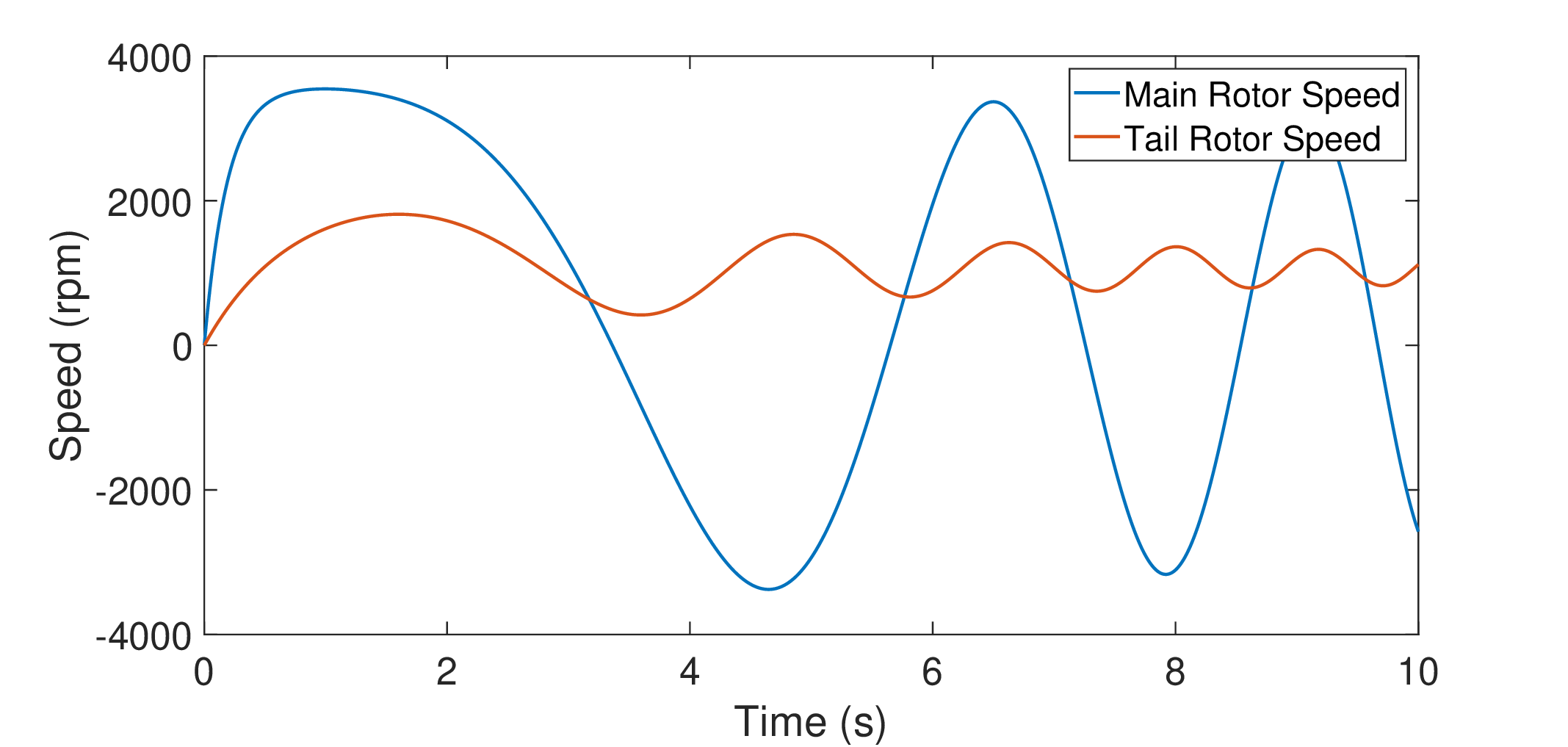}
    \caption{Bi-rotor Training Data showing showing the azimuth and pitch motors rpm}
    \label{fig:Trainig2}
\end{figure}

\begin{figure*}[htbp]
    \centering
    % First subfigure
    \begin{subfigure}[b]{0.45\textwidth}
        \centering
        \includegraphics[width=\textwidth]{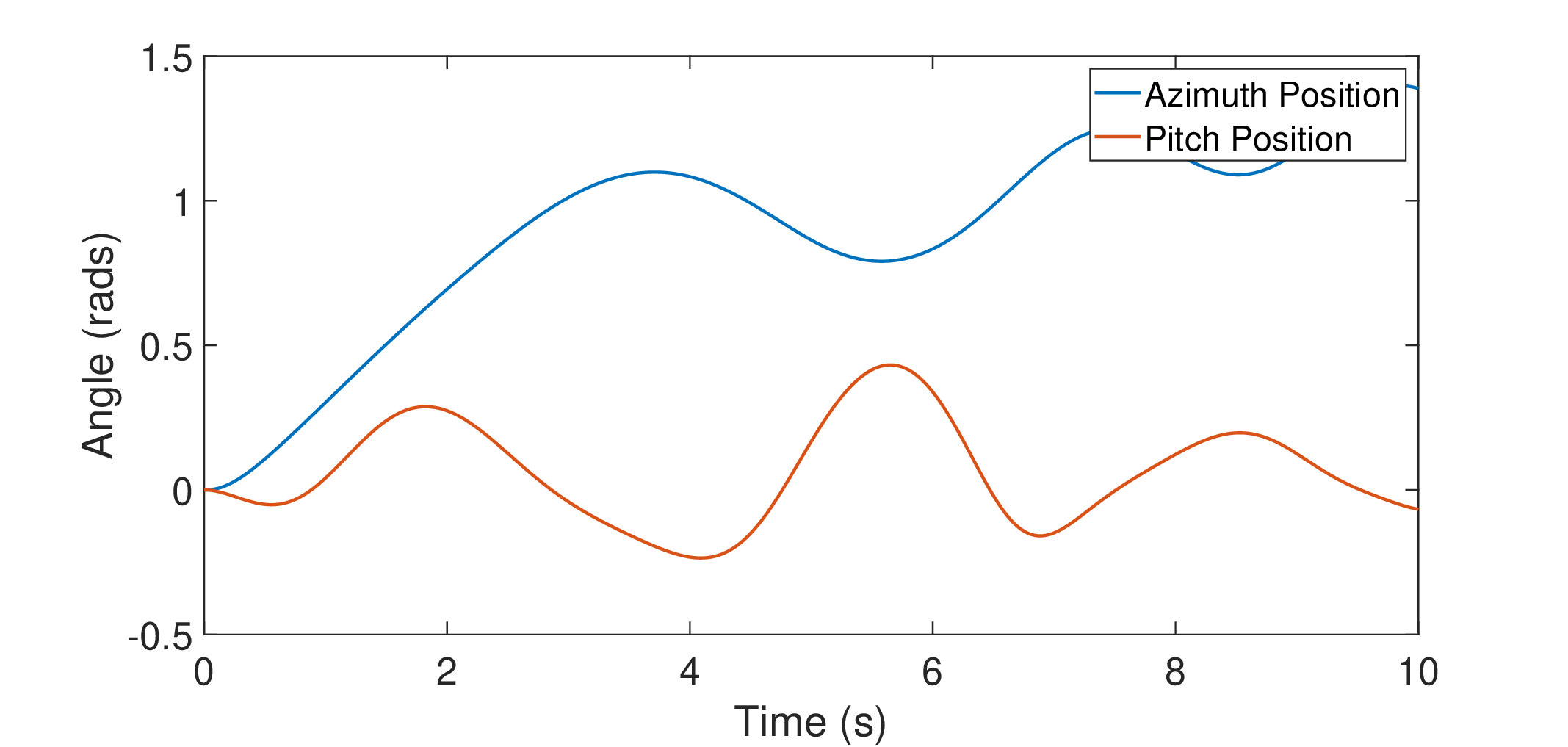}
        \caption{Azimuth and Pitch Positions}
    \end{subfigure}
    \hfill
    % Second subfigure
    \begin{subfigure}[b]{0.45\textwidth}
        \centering
        \includegraphics[width=\textwidth]{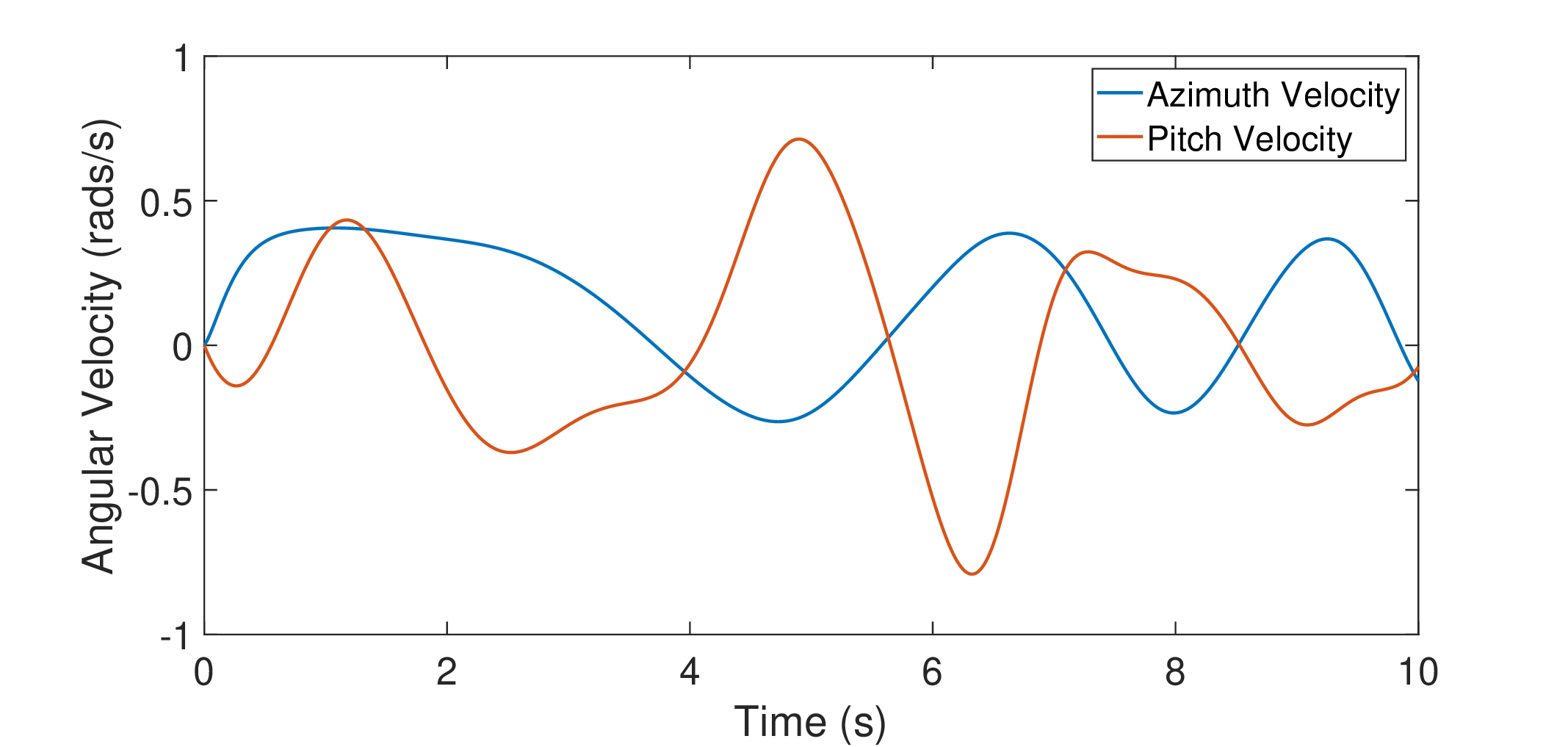}
        \caption{Azimuth and Pitch Velocities}
    
    \end{subfigure}
\caption{Bi-rotor Training Data showing showing the azimuth and pitch positions and velocities}
\label{fig:Trainig1}
\end{figure*}

The training data for the bi-rotor shown in Fig. \ref{fig:Trainig2} and \ref{fig:Trainig1} was collected over a 10-second period with a time step of 0.01 seconds. This sampling rate allows for effective capturing of the system's dynamics while balancing data resolution and computational efficiency. During the 10-second interval, a total of 1,000 data points were collected, providing a sufficient data set for training the SINDy model. This data set effectively captures the essential behaviors of the system, ensuring a more accurate and robust model identification.

\subsubsection{Sparsity Coefficient and Library Selection}

For this system, the chosen value for the sparsity coefficient $\lambda$ was 0.9 after iteratively testing values between 0 and 1.

For the library functions selection, the following candidate functions were selected as shown in the following equation which consists of polynomials, sines, cosines and combinations of them in order to accurately capture the system dynamics.

\begin{equation} 
\Theta(X, U) = 
     \begin{bmatrix}
        x_1 & ... & ... & x_6 & u \\ \\
        x_1\cos{x_2} & ... & ... & x_6\cos{x_2} & u\cos{x_2} \\ \\
        x_1^{2}\sin{x_2} & ... & ... & x_6^{2}\sin{x_2} & u^{2}\sin{x_2}\\ \\
        \sin{x_1} &... & ... & \sin{x_6} & \sin{u} \\ \\
        \cos{x_1} &... & ... & \cos{x_6} & \cos{u} 
    \end{bmatrix}  
\end{equation}

\subsubsection{TD3 Trajectory Tracking Agent}

After collecting data for model learning, it is now time to construct the TD3 agent responsible for stabilization and trajectory tracking of the bi-rotor positions.

The structure of the TD3 trajectory tracking agent is designed using four different layers as follows. First, an input layer is created consisting of 6 states representing the state vector of the agent $S_{tt}$ as follows.

\begin{equation}
    S_{tt} = 
    \begin{bmatrix}
         \alpha_a & \Omega_a & \omega_a & \alpha_p & \Omega_p & \omega_p 
    \end{bmatrix}  
\end{equation}

After that, two hidden layers were implemented using the LeakyReLU non-linear activation functions with each having 400 and 300 nodes respectively. Also, for the output layer, a Tanh activation function is implemented where the output is the input voltage used to rotate the main and tail rotors of the bi-rotor and it ranges from -1 to 1 and it is linearly interpolated to the rotation speeds of both motors.

For the reward function formulation, the following function was designed to track the behavior of  the bi-rotor.

\begin{equation}
    reward = -(ae_a^2 + be_p^2) - c(\Delta\Omega_a^2 + \Delta\Omega_p^2) + d 
\end{equation}

\begin{equation}
    e = \alpha_{desired} - \alpha_{actual} 
\end{equation}

\begin{equation}
   \Delta\Omega = \Omega_{current} - \Omega_{previous}
\end{equation}

The first term in the reward function is used to penalize the the error between the actual and desired position of the bi-rotor so that the error will tend to zero. The second term is used to penalize the oscillations in the velocity of the bi-rotor by calculating the error between the current and the previous velocity to ensure it tends to zero as without it there was excessive oscillations in tracking the trajectory. The last term is used to give a reward to the agent if the error was less than 0.01. Finally, a, b and c are constants that are weighted using trial and error to give the best performance.

The training objective was to train the bi-rotor to track a desired trajectory. The bi-rotor starts from the same position [0,0] each episode and then trains to follow a random step trajectory each episode. For each episode, there were 400 time steps with a time sample $t_s = 0.05$ for an episode length of 20 seconds. The training parameters of the agent is shown in table \ref{table:hyperparam}.

\begin{table}
\caption{Agent Training Hyperparameters}
\label{table:hyperparam}
%\begin{center}
\begin{tabular}[width = \textwidth]{|l|c|c|}
\hline
Parameter & Symbol  & Value \\ \hline
Sampling Time & $t_s$  &  0.05 seconds\\ \hline
Maximum Steps per episode & - & 400 \\ \hline
Critic Learning Rate & $\alpha$ & $1 * 10^{-4}$ \\ \hline
Actor Learning Rate & $\alpha$ & $1 * 10^{-4}$ \\ \hline
Discount Factor & $\gamma$ & 0.995\\ \hline
Buffer Size & $D$ & $2 * 10^{6}$ \\ \hline
Batch Size & $K$ & 512 \\ \hline
Noise & $\sigma$ & 0.15 \\ \hline
Noise Decay Rate & - & $1 * 10^{-3}$ \\ \hline
Target Smooth Factor & $\tau$ & 0.005 \\ \hline
Target Update Frequency & - & 10 \\ \hline
\end{tabular}
%\end{center}
\end{table}

\section{Results and Discussion}

In this section, the results of the proposed method SINDy-TD3 will be presented after applying it to the bi-rotor case study. In the beginning, the results of the SINDy algorithm will be presented to see how it captures the system dynamics accurately in order to be used as a surrogate environment to train the TD3 agent. After that, the results of training the agent will be discussed, highlighting the bi-rotor's performance in three different scenarios. The first one will be stabilizing the bi-rotor into different azimuth and pitch angles, the second and third one will be trajectory tracking with two different trajectories, sine wave trajectory and a square wave trajectory. Finally, all results will be compared to a bench mark of using TD3 only without using SINDy to highlight the importance of combining SINDy to deep reinforcement learning.

\subsection{Training SINDy Model}

\begin{figure}[H]
    \centering
    \includegraphics[width=0.7\linewidth]{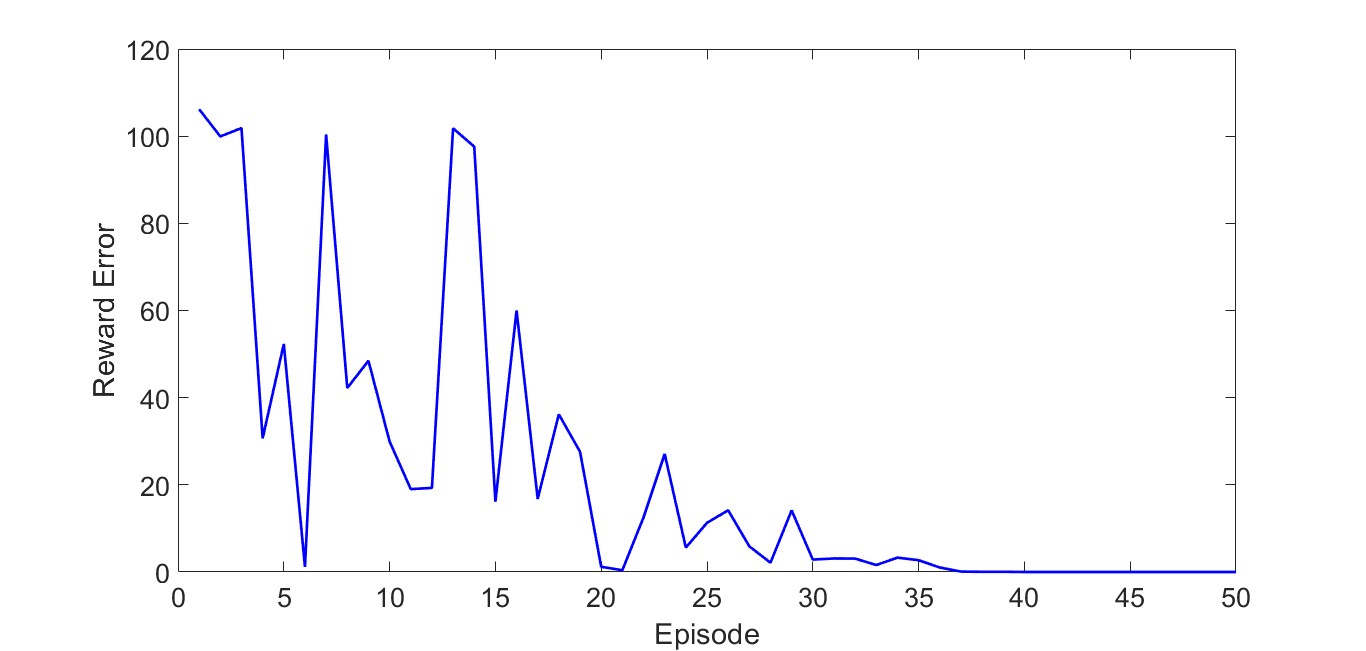}
    \caption{SINDy Model Learning Progress}
    \label{fig:SINDY-Training}
\end{figure}

To assess the accuracy of the SINDy model, the reward error between the predictions of the SINDy model and the real environment across TD3 training episodes were tracked. As shown in Fig.\ref{fig:SINDY-Training}, the reward error was initially high, indicating that the untrained SINDy model was unable to accurately capture the system dynamics. However, as the TD3 agent collected more real experience, the SINDy model was continuously updated and retrained using the newly observed data. This iterative refinement led to a noticeable decrease in reward error over time. Eventually, the SINDy model's predictions aligned closely with the real environment, making it sufficiently accurate to be used for planning and synthetic experience generation within the TD3 control loop.

\begin{figure*}[htbp]
    \centering
    % First subfigure
    \begin{subfigure}[b]{0.45\textwidth}
        \centering
        \includegraphics[width=\textwidth]{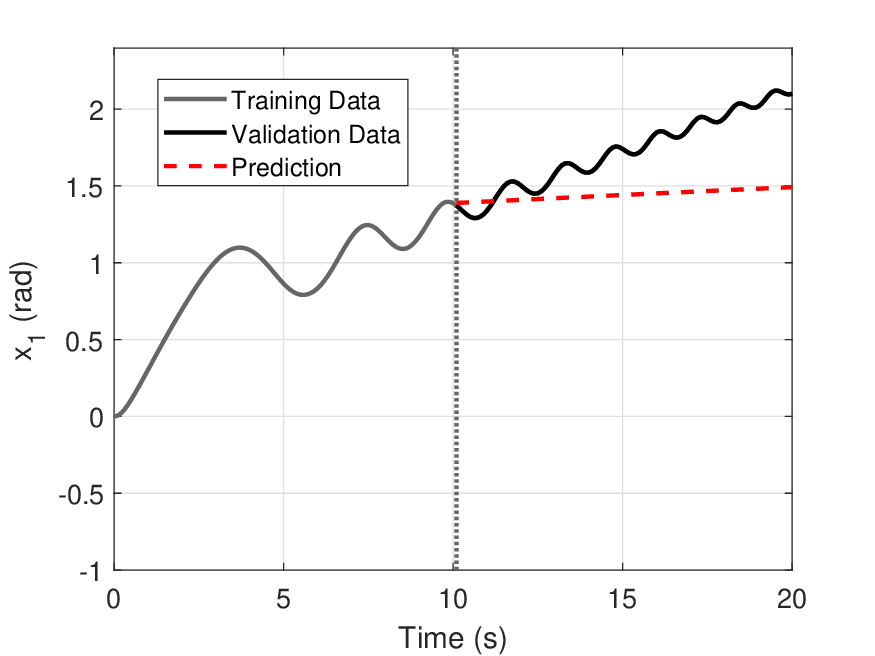}
        \caption{Azimuth position}
        \label{fig:state1bef}
    \end{subfigure}
    \hfill
    % Second subfigure
    \begin{subfigure}[b]{0.45\textwidth}
        \centering
        \includegraphics[width=\textwidth]{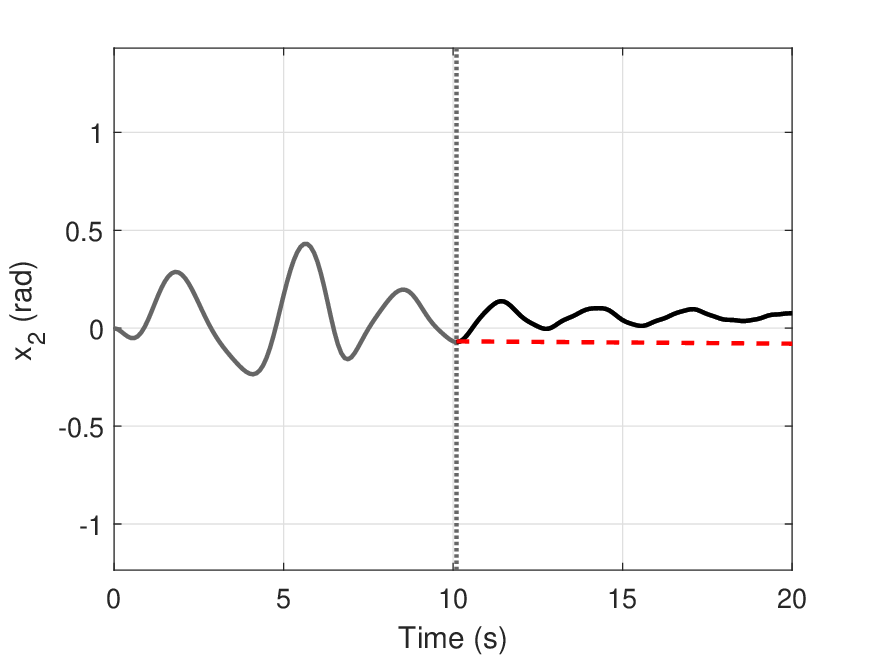}
        \caption{pitch position}
        \label{fig:state2bef}
    \end{subfigure}
    \vspace{0.5cm}
    % Third subfigure
    \begin{subfigure}[b]{0.45\textwidth}
        \centering
        \includegraphics[width=\textwidth]{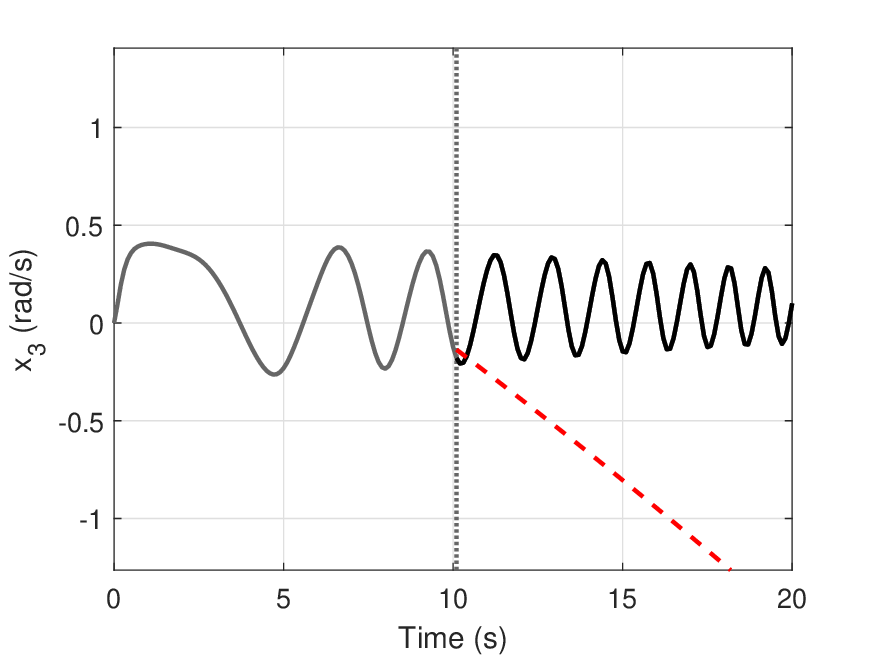}
        \caption{Azimuth angular velocity}
        \label{fig:state3bef}
    \end{subfigure}
    \hfill
    % Fourth subfigure
    \begin{subfigure}[b]{0.45\textwidth}
        \centering
        \includegraphics[width=\textwidth]{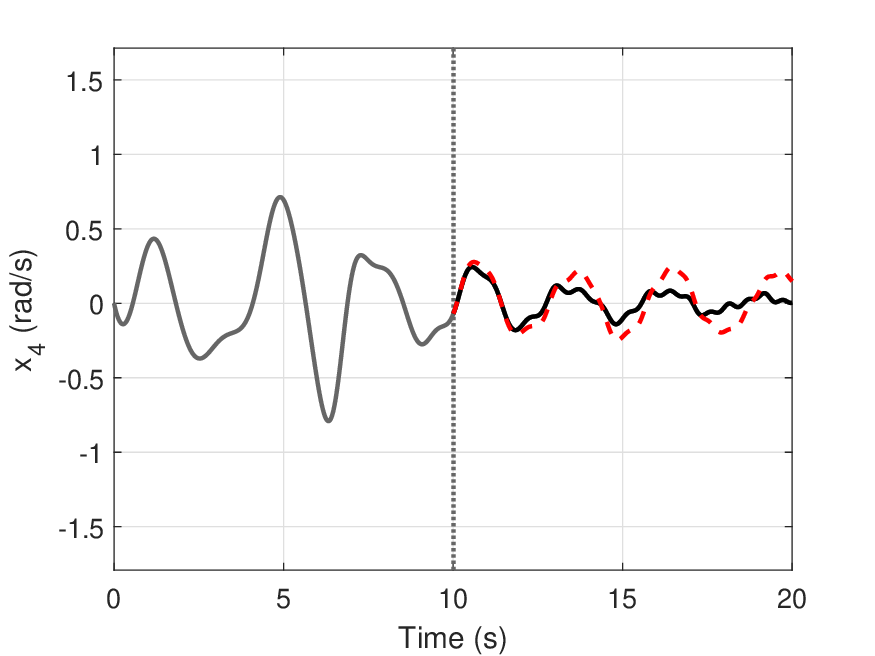}
        \caption{pitch angular velocity}
        \label{fig:state4bef}
    \end{subfigure}
    \vspace{0.5cm}
    % Fifth subfigure
    \begin{subfigure}[b]{0.45\textwidth}
        \centering
        \includegraphics[width=\textwidth]{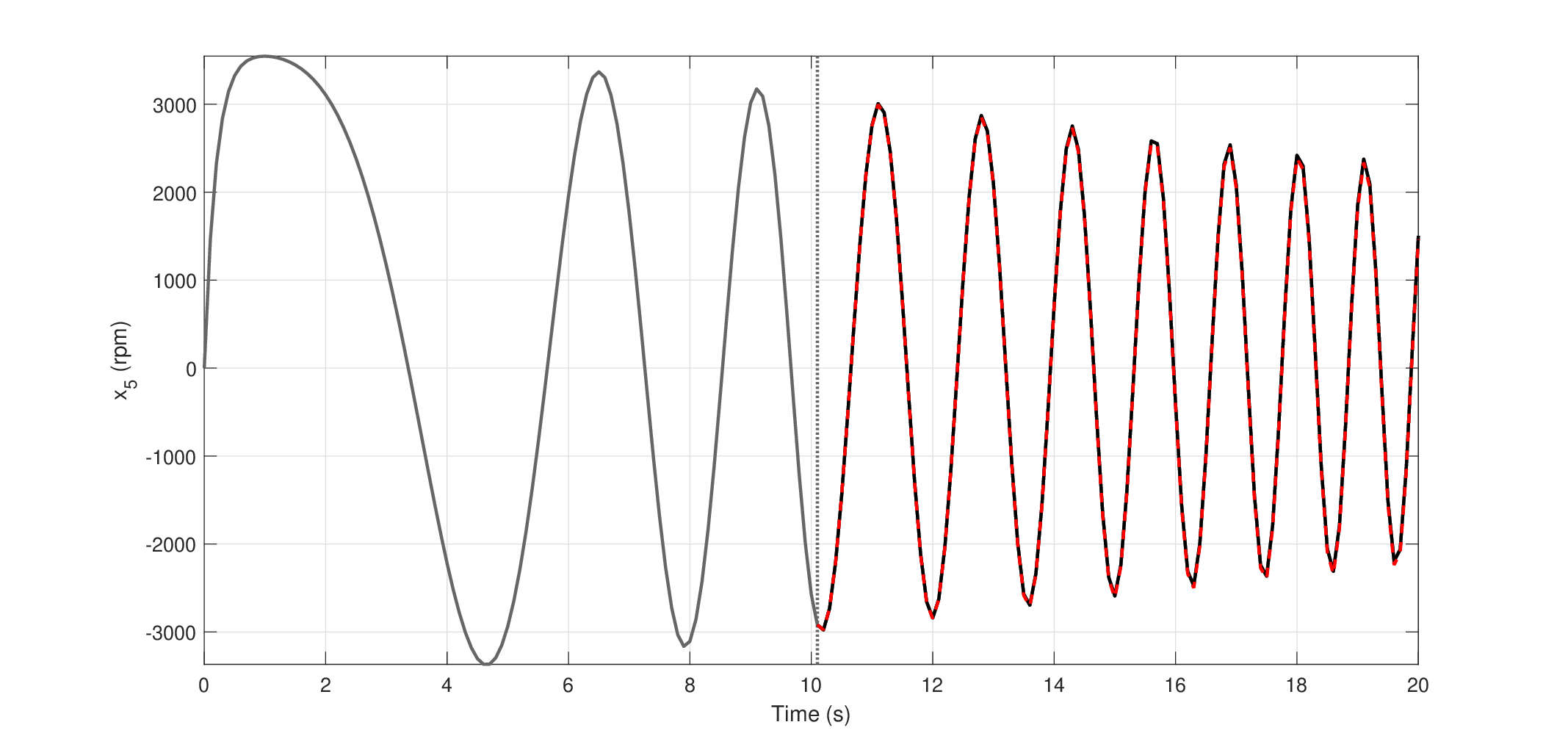}
        \caption{Main rotor angular velocity}
        \label{fig:state5bef}
    \end{subfigure}
    \hfill
    % Sixth subfigure
    \begin{subfigure}[b]{0.45\textwidth}
        \centering
        \includegraphics[width=\textwidth]{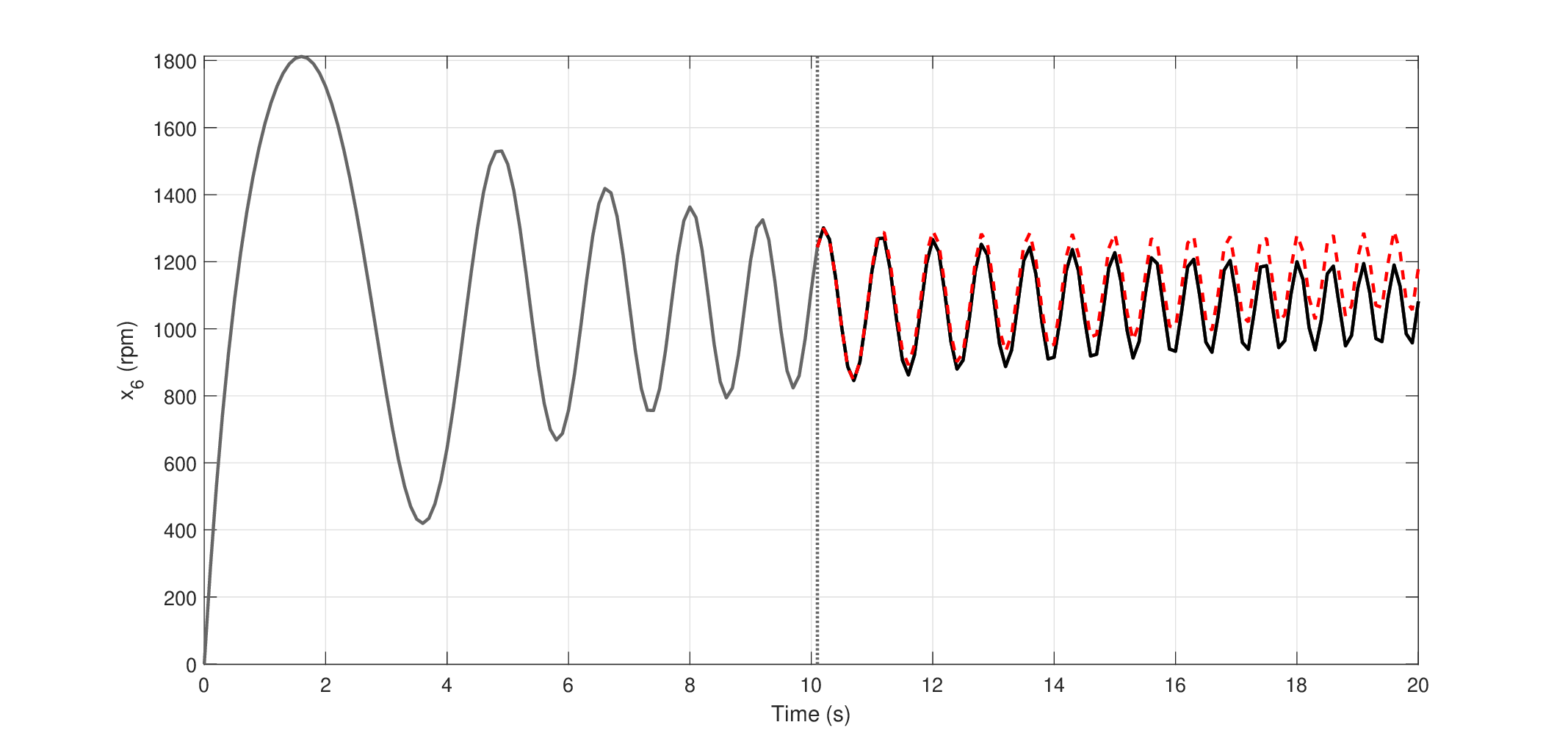}
        \caption{Tail rotor angular velocity}
        \label{fig:state6bef}
    \end{subfigure}
    
    \caption{Comparison between validation data and the SINDy model’s prediction before reinforcement learning training.}
    \label{fig:all_states_before}
\end{figure*}

Figure \ref{fig:all_states_before} displays the performance of the identified SINDy model before the implementation of reinforcement learning. The model was only trained on the initial 10 seconds of a chirp signal trajectory. The subplots labeled (a) through (f) show the progression of six system states: the azimuth and pitch positions, their corresponding angular velocities, and the angular velocities of both the main and tail rotors. While the model effectively captures the dynamics of the rotor angular velocities with minimal errors, it has difficulty accurately representing the azimuth and pitch positions and velocities. It is anticipated that as more data is gathered from the trajectories during TD3 training, the model's precision will improve considerably for all states.

To assess the accuracy of the SINDy model after it was trained during reinforcement learning, a comparison was conducted between the SINDy-predicted dynamics and the real environment. The same chirp signal used during the initial training phase was applied as a forcing input to both the final identified SINDy model and the real birotor environment.

\begin{figure*}[htbp]
    \centering
    % First subfigure
    \begin{subfigure}[b]{0.45\textwidth}
        \centering
        \includegraphics[width=\textwidth]{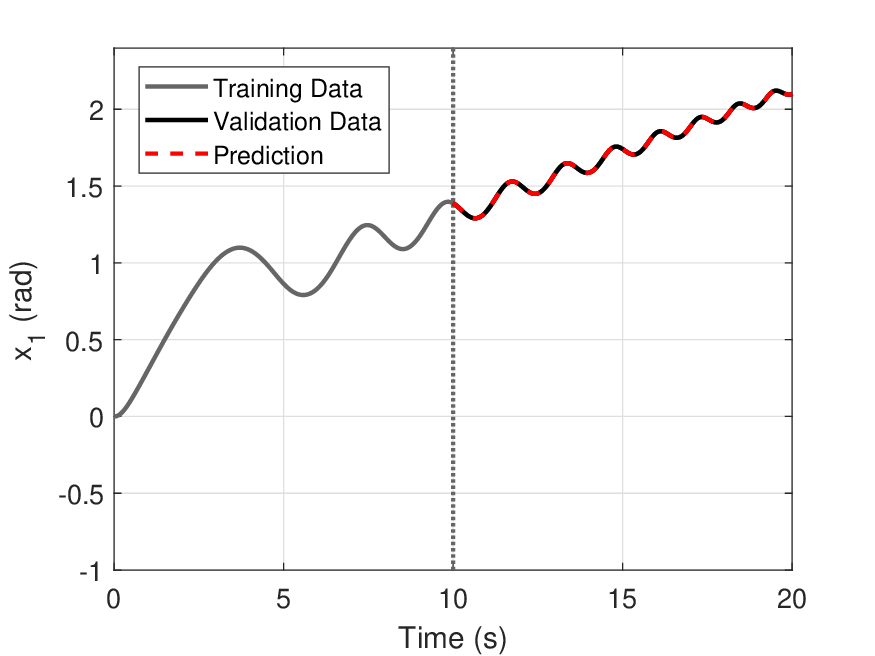}
        \caption{Azimuth position}
        \label{fig:state1}
    \end{subfigure}
    \hfill
    % Second subfigure
    \begin{subfigure}[b]{0.45\textwidth}
        \centering
        \includegraphics[width=\textwidth]{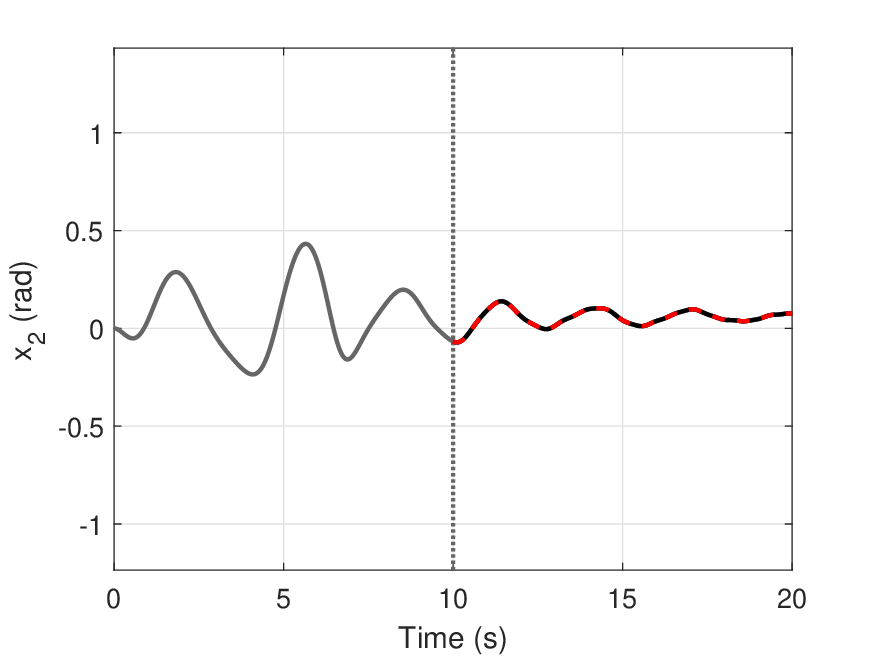}
        \caption{pitch position}
        \label{fig:state2}
    \end{subfigure}
    \vspace{0.5cm}
    % Third subfigure
    \begin{subfigure}[b]{0.45\textwidth}
        \centering
        \includegraphics[width=\textwidth]{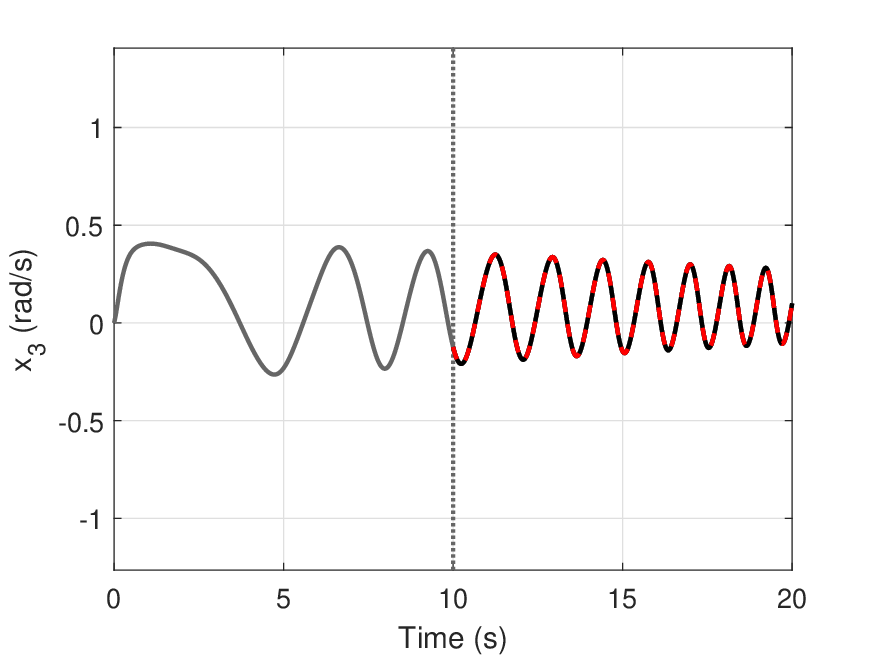}
        \caption{Azimuth angular velocity}
        \label{fig:state3}
    \end{subfigure}
    \hfill
    % Fourth subfigure
    \begin{subfigure}[b]{0.45\textwidth}
        \centering
        \includegraphics[width=\textwidth]{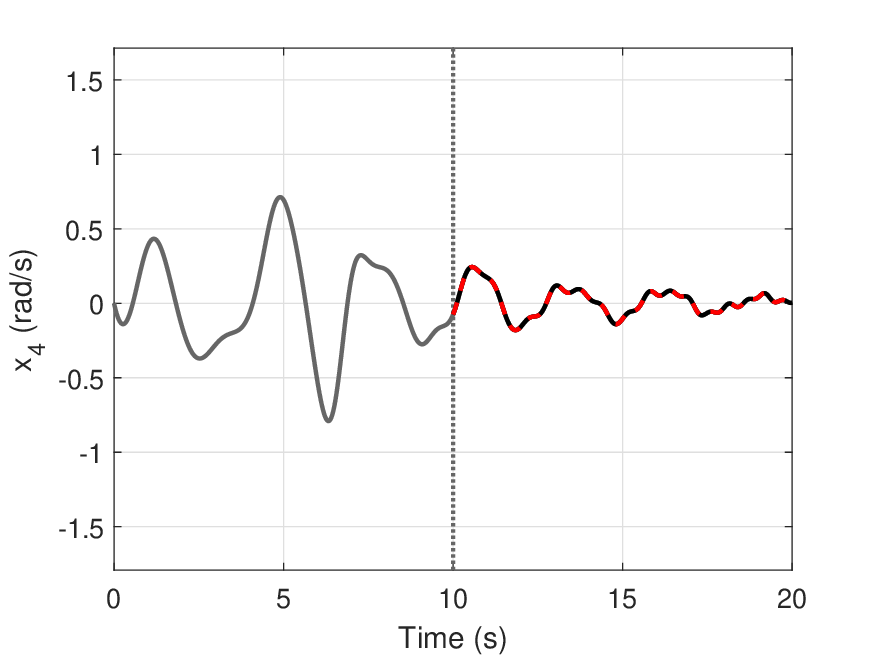}
        \caption{pitch angular velocity}
        \label{fig:state4}
    \end{subfigure}
    \vspace{0.5cm}
    % Fifth subfigure
    \begin{subfigure}[b]{0.45\textwidth}
        \centering
        \includegraphics[width=\textwidth]{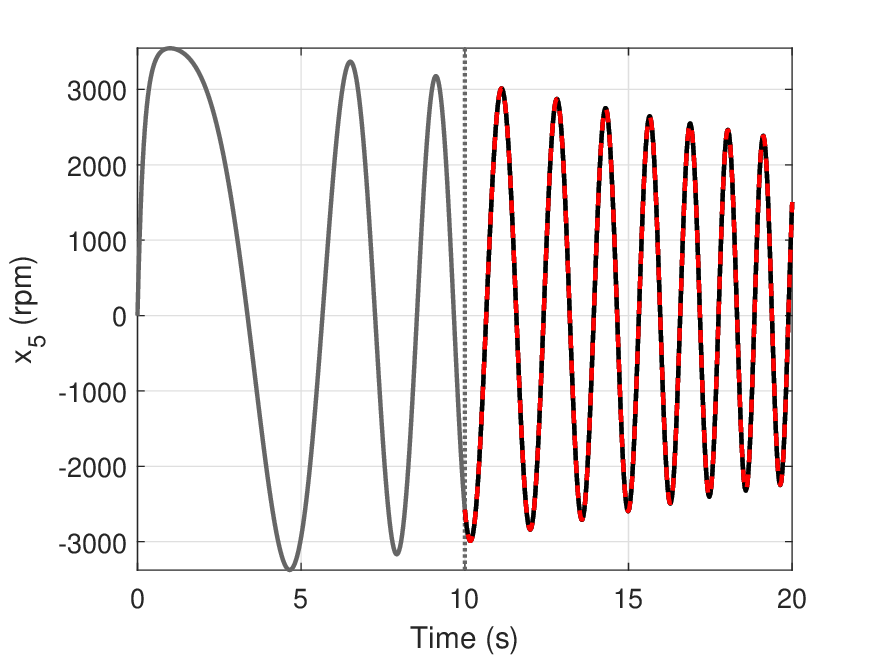}
        \caption{Main rotor angular velocity}
        \label{fig:state5}
    \end{subfigure}
    \hfill
    % Sixth subfigure
    \begin{subfigure}[b]{0.45\textwidth}
        \centering
        \includegraphics[width=\textwidth]{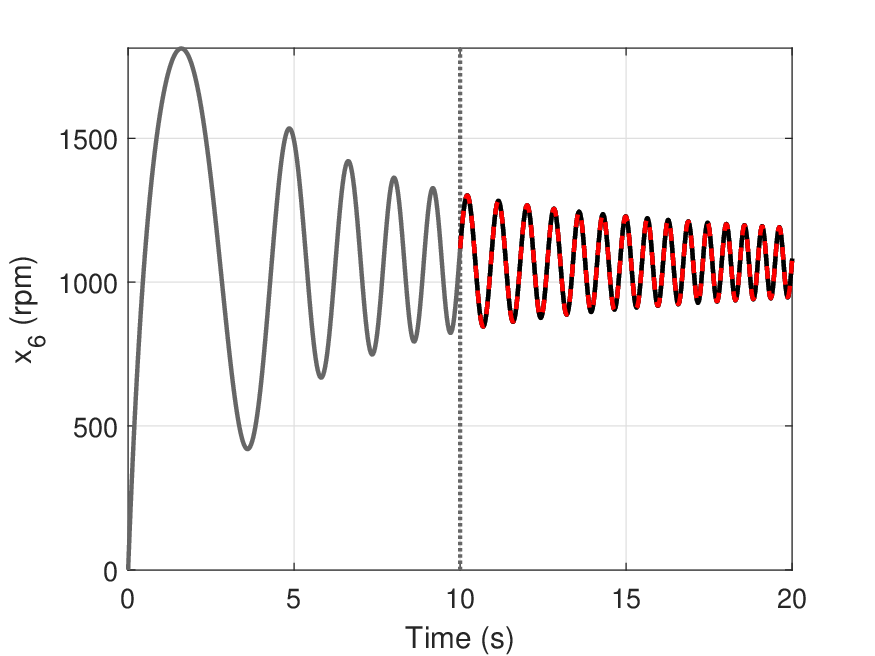}
        \caption{Tail rotor angular velocity}
        \label{fig:state6}
    \end{subfigure}
    
    \caption{Comparison between validation data and the SINDy model’s prediction after reinforcement learning training.}
    \label{fig:all_states}
\end{figure*}

Fig.~\ref{fig:all_states} shows the resulting trajectories, where the real environment's response is represented by the black line and the SINDy model's prediction is shown by the dotted red line. As observed in the figure, the trained SINDy model closely replicates the true system behavior, with minimal deviation between the predicted and actual trajectories. This demonstrates that, after being iteratively refined throughout reinforcement learning, the SINDy model becomes sufficiently accurate to serve as a reliable simulator for planning and experience generation.

Table \ref{tab:model} shows the $\Xi$ matrix which shows the sparse elements of each state of the dynamics of the bi-rotor. After evaluating $\Dot{X}$ we will reach the following dynamics for the bi-rotor.

\begin{table*}[htbp]
    
\centering
\caption{SINDY model $\Xi$ matrix}
\label{tab:model}
\begin{tabular}{|c|c|c|c|c|c|c|}
\hline
Library Function & $\Dot{x}_1$ & $\Dot{x}_2$ & $\Dot{x}_3$ & $\Dot{x}_4$ & $\Dot{x}_5$ & $\Dot{x}_6$ \\ \hline
1                 &   0       &    0      &  0                     & 0.5961    & 262.7             & -8.54         \\ \hline
$x_1$             &   0       &    0      &  0                     & 0         & 0                 & 0        \\ \hline
$x_2$             &   0       &    0      &  0                     & 0         & 0                 & 0        \\ \hline
$x_3$             &   1       &    0      &  -10.0573              & 0         & 0                 & 0        \\ \hline
$x_4$             &   0       &    1      &  0                     & -0.414    & 0                 & 0        \\ \hline
$x_5$             &   0       &    0      &  0                     & 0         & -5.2281           & 0        \\ \hline
$x_6$             &   0       &    0      &  0                     & 0.0013    & 0                 & -1.4193  \\ \hline
$u_1$             &   0       &    0      &  0                     & 0.1303    & $3.7 \times 10^4$ & 0        \\ \hline
$u_2$             &   0       &    0      &  0                     & 0         & 0         & $6.1 \times 10^3$\\ \hline
$x_1 \cos{x_2}$   &   0       &    0      &  0                     & 0         & 0         & 0         \\ \hline
$x_2 \cos{x_2}$   &   0       &    0      &  0                     & 0         & 0         & 0         \\ \hline
$x_3 \cos{x_2}$   &   0       &    0      &  0                     & 0         & 0         & 0         \\ \hline
$x_4 \cos{x_2}$   &   0       &    0      &  0                     & 0         & 0         & 0         \\ \hline
$x_5 \cos{x_2}$   &   0       &    0      &  $9.8316\times10^{-4}$ & 0         & 0         & 0         \\ \hline
$x_6 \cos{x_2}$   &   0       &    0      &  0                     & 0         & 0         & 0         \\ \hline
$u_1 \cos{x_2}$   &   0       &    0      &  0                     & 0         & 0         & 0         \\ \hline
$u_2 \cos{x_2}$   &   0       &    0      &  -0.7025               & 0         & 0         & 0         \\ \hline
$x_1^2 \sin{x_2}$ &   0       &    0      &  0                     & 0         & 0         & 0         \\ \hline
$x_2^2 \sin{x_2}$ &   0       &    0      &  0                     & 0         & 0         & 0         \\ \hline
$x_3^2 \sin{x_2}$ &   0       &    0      &  0                     & -16.3547  & 0         & 0         \\ \hline
$x_4^2 \sin{x_2}$ &   0       &    0      &  0                     & 0         & 0         & 0         \\ \hline
$x_5^2 \sin{x_2}$ &   0       &    0      &  0                     & 0         & 0         & 0         \\ \hline
$x_6^2 \sin{x_2}$ &   0       &    0      &  0                     & 0         & 0         & 0         \\ \hline
$u_1^2 \sin{x_2}$ &   0       &    0      &  0                     & 0         & 0         & 0         \\ \hline
$u_2^2 \sin{x_2}$ &   0       &    0      &  0                     & 0         & 0         & 0         \\ \hline
$\sin{x_2}$       &   0       &    0      &  0                     & -3.54373  & 0         & 0         \\ \hline
$\cos{x_2}$       &   0       &    0      &  0.9479                & -1.7473   & 0         & 0         \\ \hline
\end{tabular}

\end{table*}

After seeing the dynamics shown in Table \ref{tab:model} that was obtained by the SINDY model   and comparing it with the real environment of the system, we can infer that the SINDy model successfully learns the underlying system dynamics with a minimal amount of training data. The excellent agreement observed during the prediction phase indicates the robustness of the model in generalizing beyond the training data. This finding reinforces the effectiveness of using sparse identification techniques, like SINDy, for modeling complex, nonlinear systems such as the bi-rotor. \\ \\ \\

\subsection{Reinforcement Learning Control}

\begin{figure}[H]
    \centering
    \includegraphics[width=0.8\linewidth]{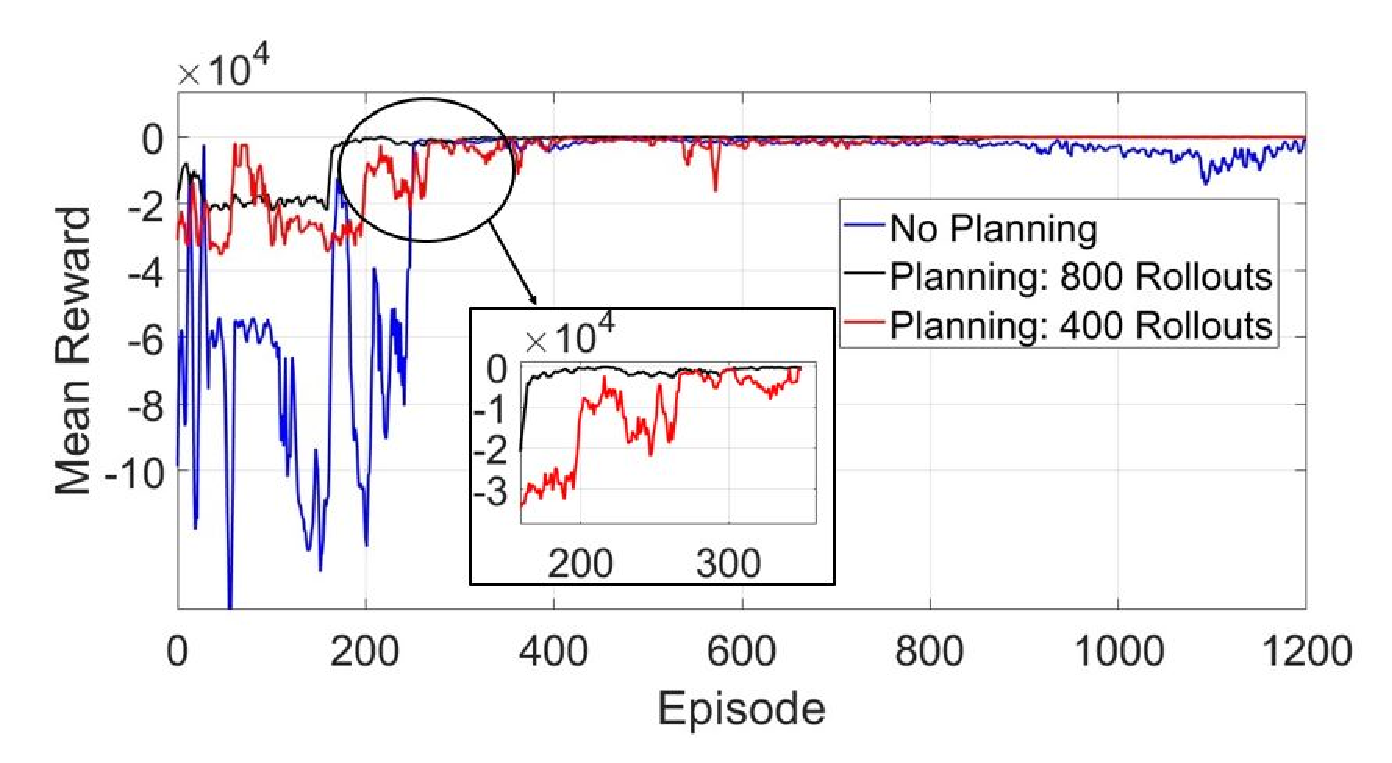}
    \caption{Dyna-Style vs. Direct TD3 Average Rewards}
    \label{fig:RL-Training}
\end{figure}

Fig. \ref{fig:RL-Training} illustrates the learning performance of the proposed Dyna-Style SINDy-TD3 algorithm under different planning conditions. The first case is a benchmark with no planning (just Direct-RL), as well as two different scenarios of planning where in one case 400 synthetic rollouts from the SINDy model were injected to the training buffer, and in the second case 800 synthetic rollouts were injected. As shown in the figure, in the beginning of the training the rewards were very low due to the penalty of exceeding the system bounds, then after gaining more experience the rewards starts to increase. Both planning configurations lead to a faster convergence of the average reward to reach the desired average required compared to the non-planning baseline. Also, training was more stable in both planning cases with significantly greater rewards in the early learning phase. Furthermore, increasing the rollouts from 400 to 800 increased the average reward acquired by the agent and led to faster convergence to reach the required average reward in just 180 episodes. All configurations reached nearly the same average reward in the end; however, model-based planning was significantly faster indicating that the main advantage of model-based planning lies in improving the sample efficiency where it becomes very useful in environments with low data limit. 

\subsection{Scenario A: Step Trajectory}

\begin{figure} [H]
    \centering
    \includegraphics[width=0.8\linewidth]{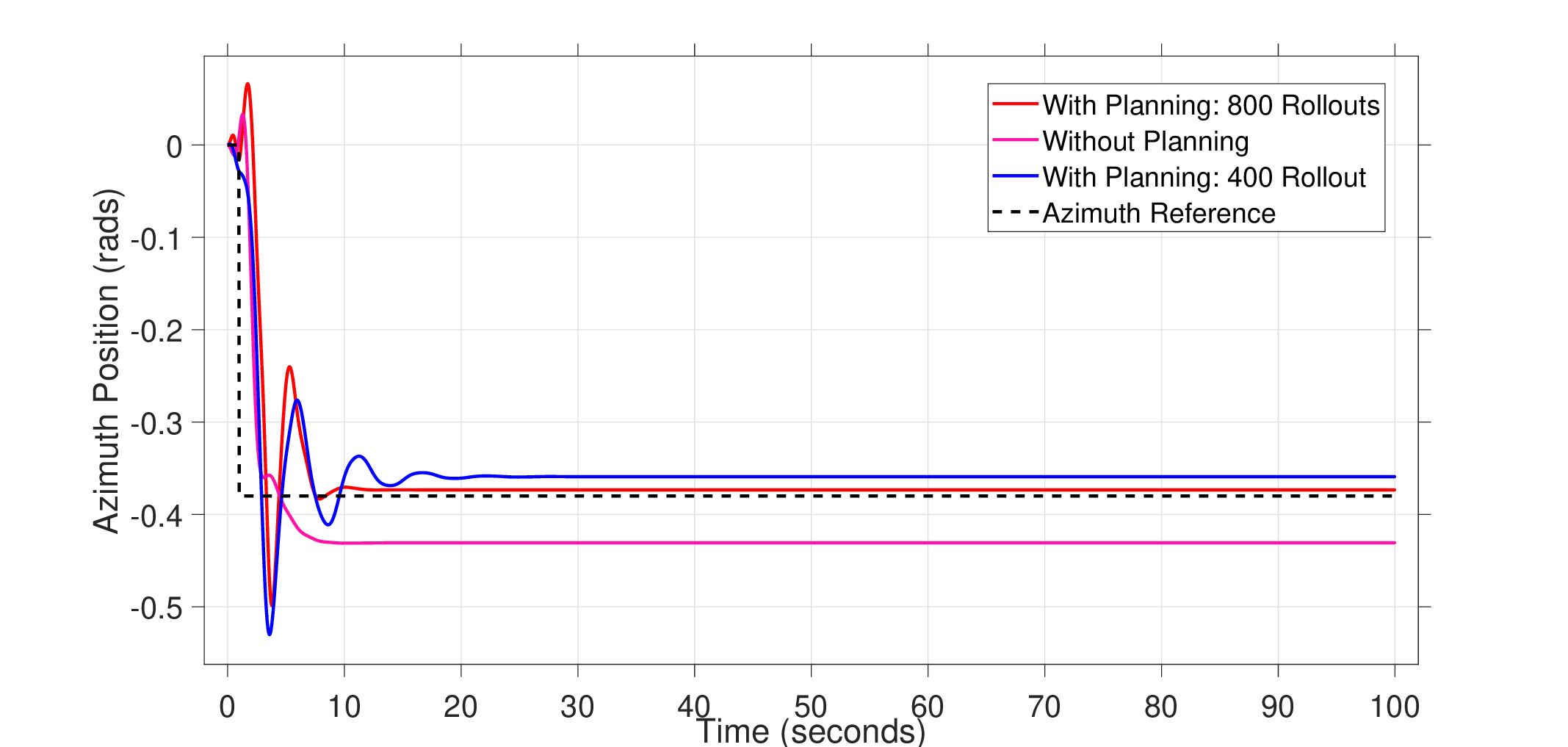}
    \caption{Azimuth Position Step Trajectory Tracking}
    \label{fig:step-azimuth}
\end{figure}

Fig. \ref{fig:step-azimuth} shows the azimuth position step trajectory tracking in three different configurations: without planning, planning with 400 rollouts and planning with 800 rollouts. In the non-planning configurations the agent was able to stabilize however far from the reference required with a steady-state error of about 0.08 rads. After injecting 400 rollout to the training buffer, the agent was able to successfully track the reference at a steady-state error better than the non-planning and with less interactions with the real environment however, with a greater overshoot. Then the 800 rollouts was tested and it shows the best tracking of the reference with minimal steady-state error and interactions with the real environment than both the 400 rollouts and the non-planning enhancing the sample efficiency of the training.

\begin{figure} [H]
    \centering
    \includegraphics[width=0.8\linewidth]{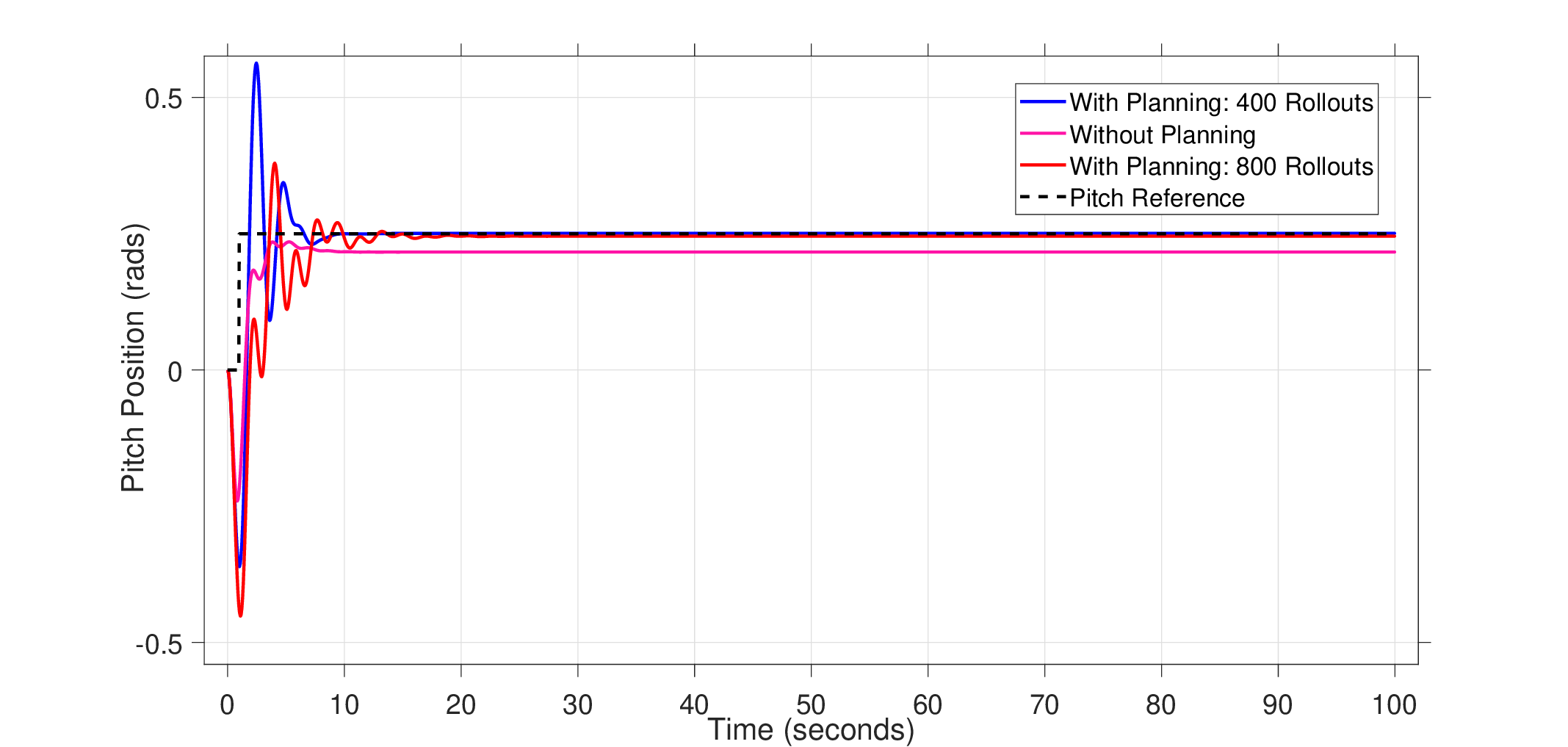}
    \caption{Pitch Position Step Trajectory Tracking}
    \label{fig:step-pitch}
\end{figure}

Fig. \ref{fig:step-pitch} shows the pitch position tracking performance under the same three configurations. The non-planning setup stabilizes with a noticeable steady-state error. Introducing 400 rollouts improves tracking accuracy but results in higher overshoot. With 800 rollouts, the agent achieves the best performance—closely tracking the reference with minimal overshoot and steady-state error—demonstrating improved learning and reduced dependency on real environment interactions.

\subsection{Scenario B: Sine Wave Trajectory}

\begin{figure}[H]
    \centering

    % --- First Row: SINDy-TD3 Tracking ---
    \begin{subfigure}{\textwidth}
        \centering
        \begin{subfigure}{0.49\textwidth}
            \centering
            \includegraphics[width=\textwidth]{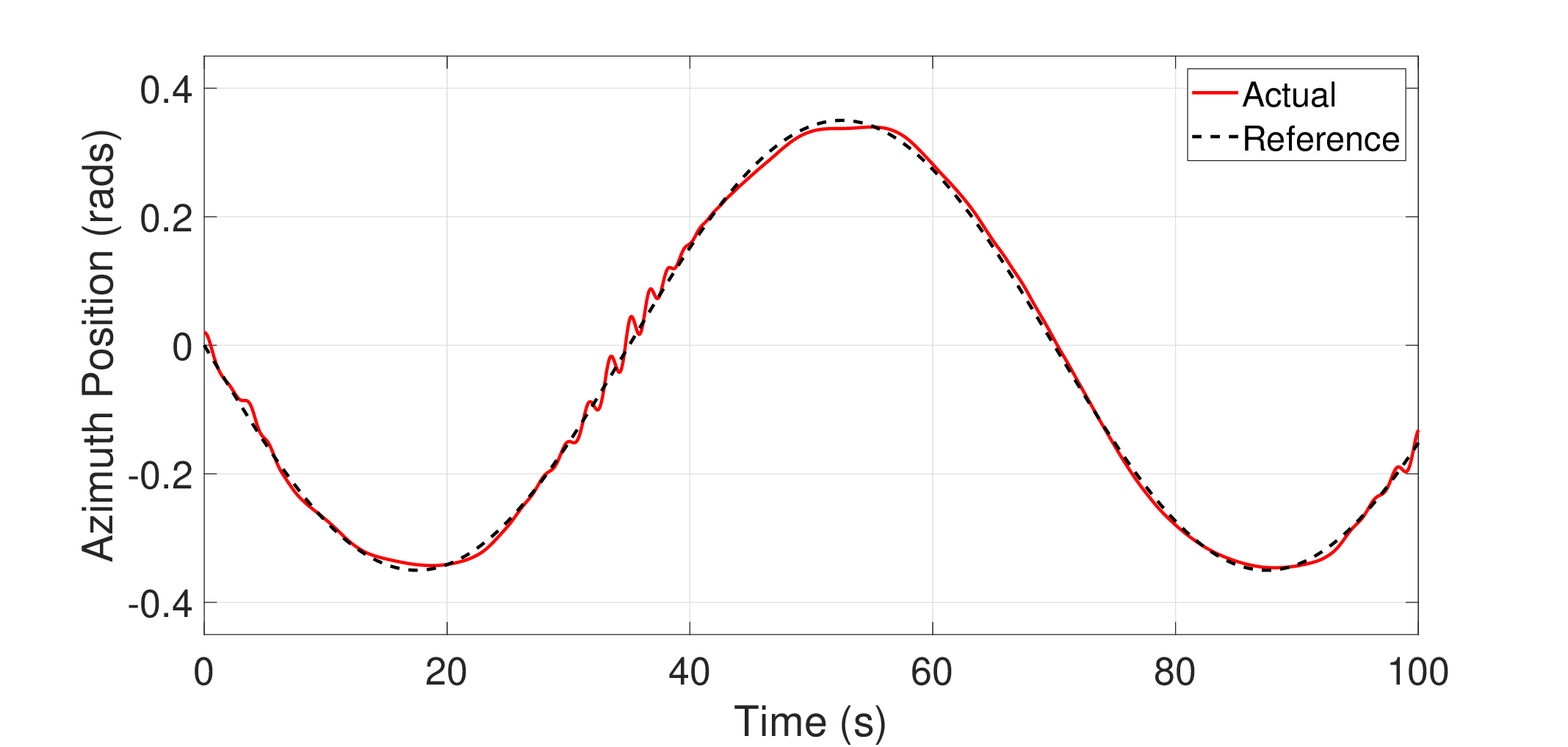}
            \caption{Azimuth position (SINDy-TD3)}
            \label{fig:azimuth-sine-sindy-td3}

        \end{subfigure}
        \hfill
        \begin{subfigure}{0.49\textwidth}
            \centering
            \includegraphics[width=\textwidth]{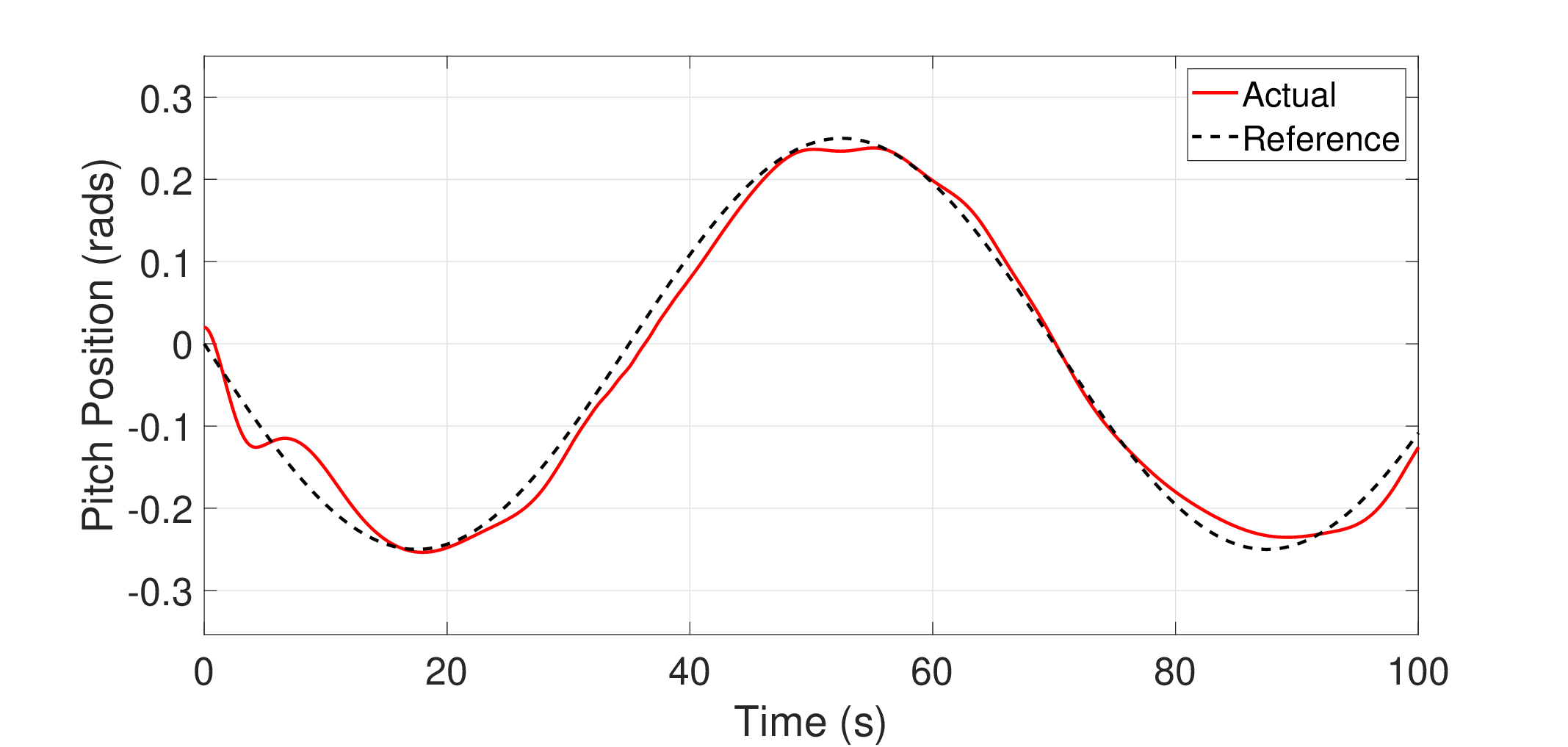}
            \caption{Pitch position (SINDy-TD3)}
        \end{subfigure}
       % \caption*{\textbf{(a)} SINDy-TD3 Sine Wave Trajectory Tracking}
        \label{fig:sine-sindy-td3}

    \end{subfigure}

    \vspace{0.2cm}

    % --- Second Row: TD3 Tracking ---
    \begin{subfigure}{\textwidth}
        \centering
        \begin{subfigure}{0.49\textwidth}
            \centering
            \includegraphics[width=\textwidth]{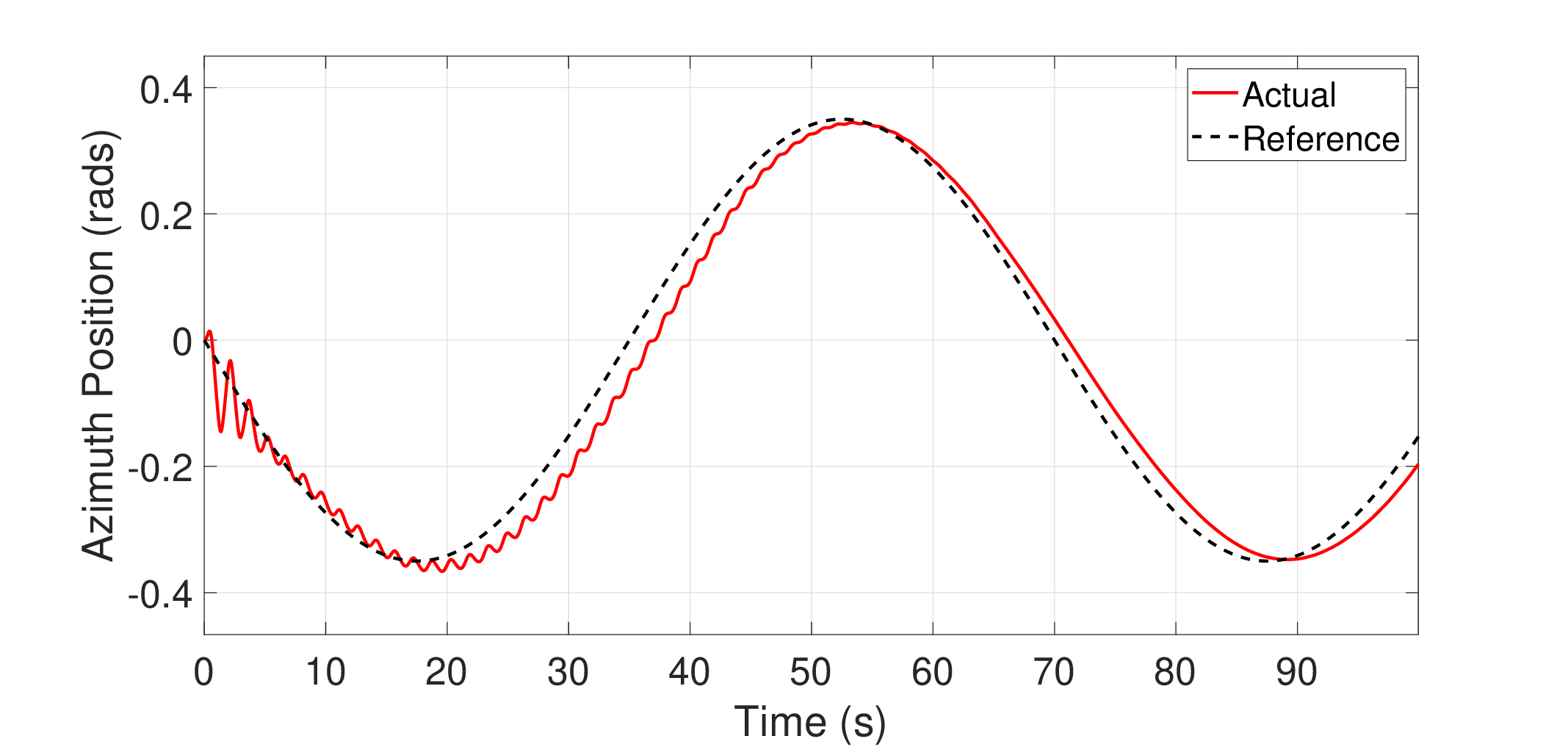}
            \caption{Azimuth position (TD3)}
        \end{subfigure}
        \hfill
        \begin{subfigure}{0.49\textwidth}
            \centering
            \includegraphics[width=\textwidth]{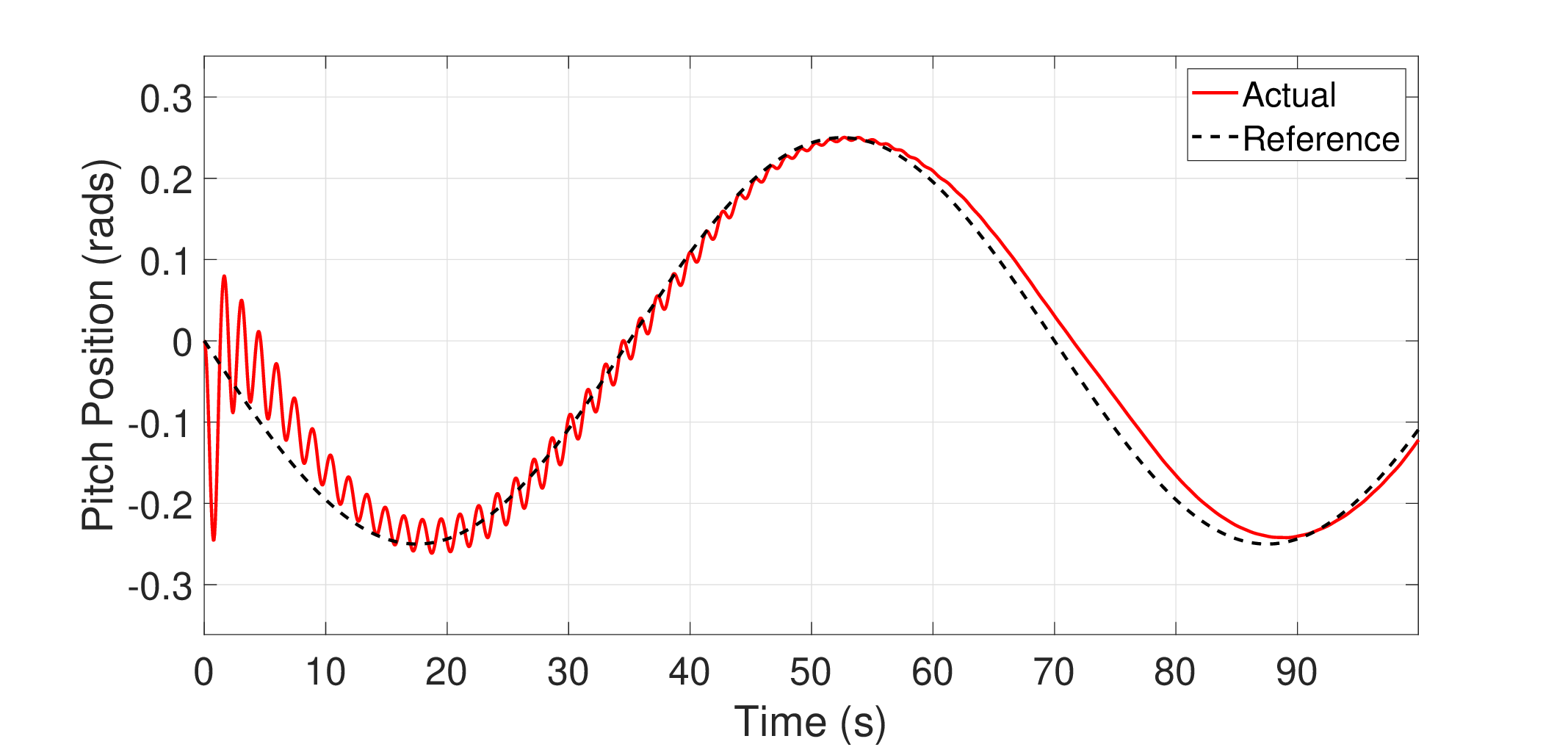}
            \caption{Pitch position (TD3)}
        \end{subfigure}
        \label{sine-td3}
    %    \caption*{\textbf{(b)} TD3 Sine Wave Trajectory Tracking}
    \end{subfigure}

    \caption{Comparison between SINDy-TD3 and standard TD3 in tracking performance for azimuth and pitch angles.}
    \label{fig:comparison-sindytd3-td3}
\end{figure}

For the sine and square wave trajectory tracking which will be presented in this section and the following one, the agent will be subjected to further 1300 episodes of training in order to accurately follow the desired trajectories. For the sine wave trajectory, the reference trajectory for the azimuth angle was chosen to be 0.35 radians amplitude and 0.015 Hz frequency and 0.25 radians amplitude for the pitch angle reference and the same frequency. The sine wave trajectory tracking results using the SINDy-TD3 agent are presented in Figure \ref{fig:comparison-sindytd3-td3} (a) and (b) for the azimuth and pitch angles respectively. Fig. 14 (a) shows that the azimuth angle closely follows the reference sine wave trajectory, with minimal steady state error. The transient response is observed primarily at the beginning, after which the agent effectively tracks the sinusoidal reference. Similarly, the pitch angle demonstrates successful tracking of the sine wave trajectory. While there are minor deviations in amplitude and phase, particularly during the initial cycles, the overall performance shows good agreement between the predicted and reference trajectories. These results underline the ability of the SINDy-TD3 agent to generalize and adapt to more complex trajectory tracking scenarios, reinforcing its robustness and reliability for dynamic control tasks.

To compare the results, the same trajectory was then done without planning, just direct TD3 without the use of SINDy to validate the accuracy of SINDy-TD3. As shown in Figure \ref{fig:comparison-sindytd3-td3} (c) and (d), the TD3 agent was able to accurately track the desired trajectory however, there are more oscillations in the response than SINDy-TD3. Also, training the agent without SINDy needed about 4500 episode in total to get those results while the SINDy-TD3 trained only 2500 episode which is more data-efficient than TD3 only.

\subsection{Scenario C: Square Wave Trajectory}

For the square wave trajectory, the reference trajectory for the azimuth angle was chosen to be 0.4 radians amplitude and 0.01 Hz frequency and -0.5 radians amplitude for the pitch angle reference and the same frequency. 

\begin{figure}[H]
    \centering

    % --- First Row: SINDy-TD3 Square Wave ---
    \begin{subfigure}{\textwidth}
        \centering
        \begin{subfigure}{0.49\textwidth}
            \centering
            \includegraphics[width=\textwidth]{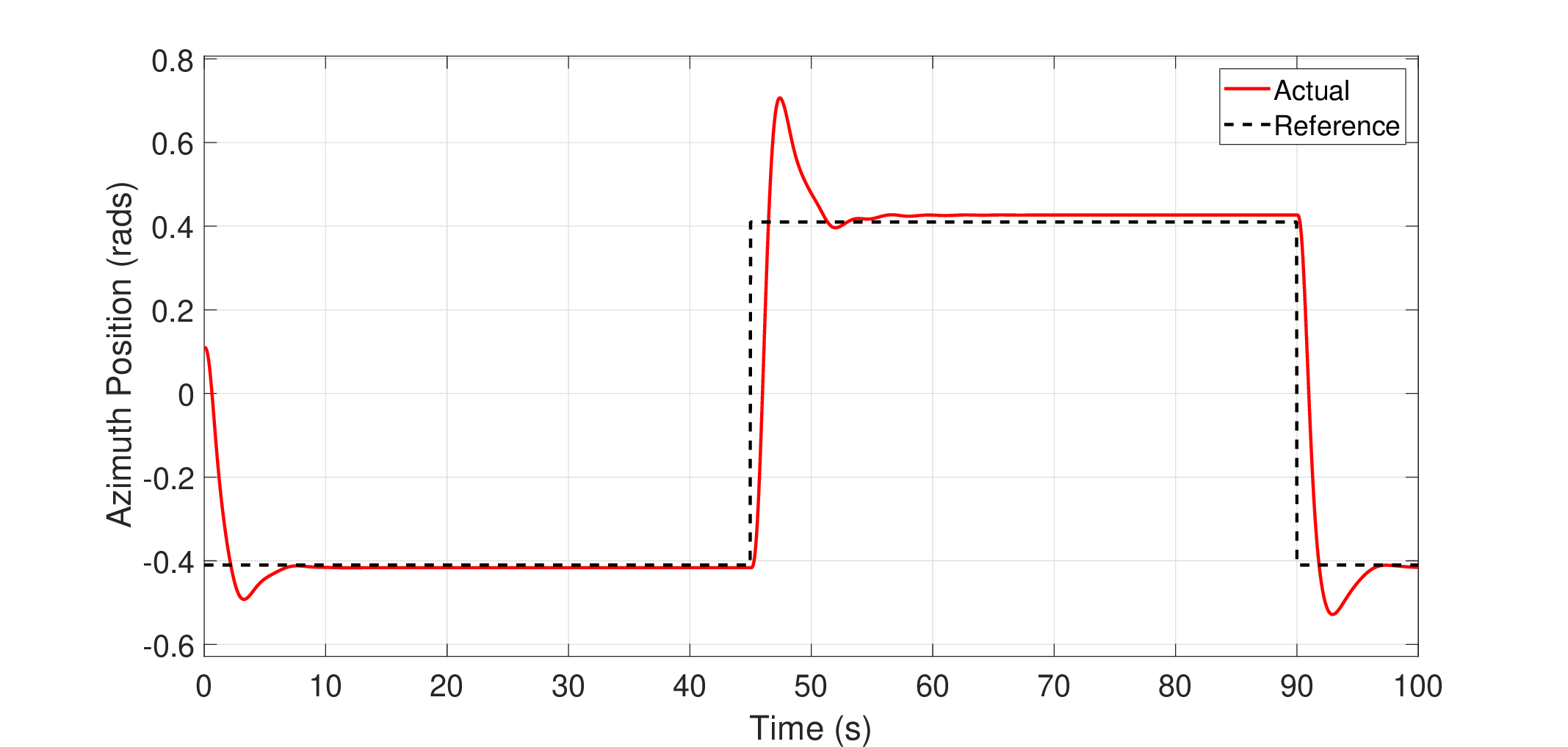}
            \caption{Azimuth position (SINDy-TD3)}
            \label{fig:sindy-square-azimuth}
        \end{subfigure}
        \hfill
        \begin{subfigure}{0.49\textwidth}
            \centering
            \includegraphics[width=\textwidth]{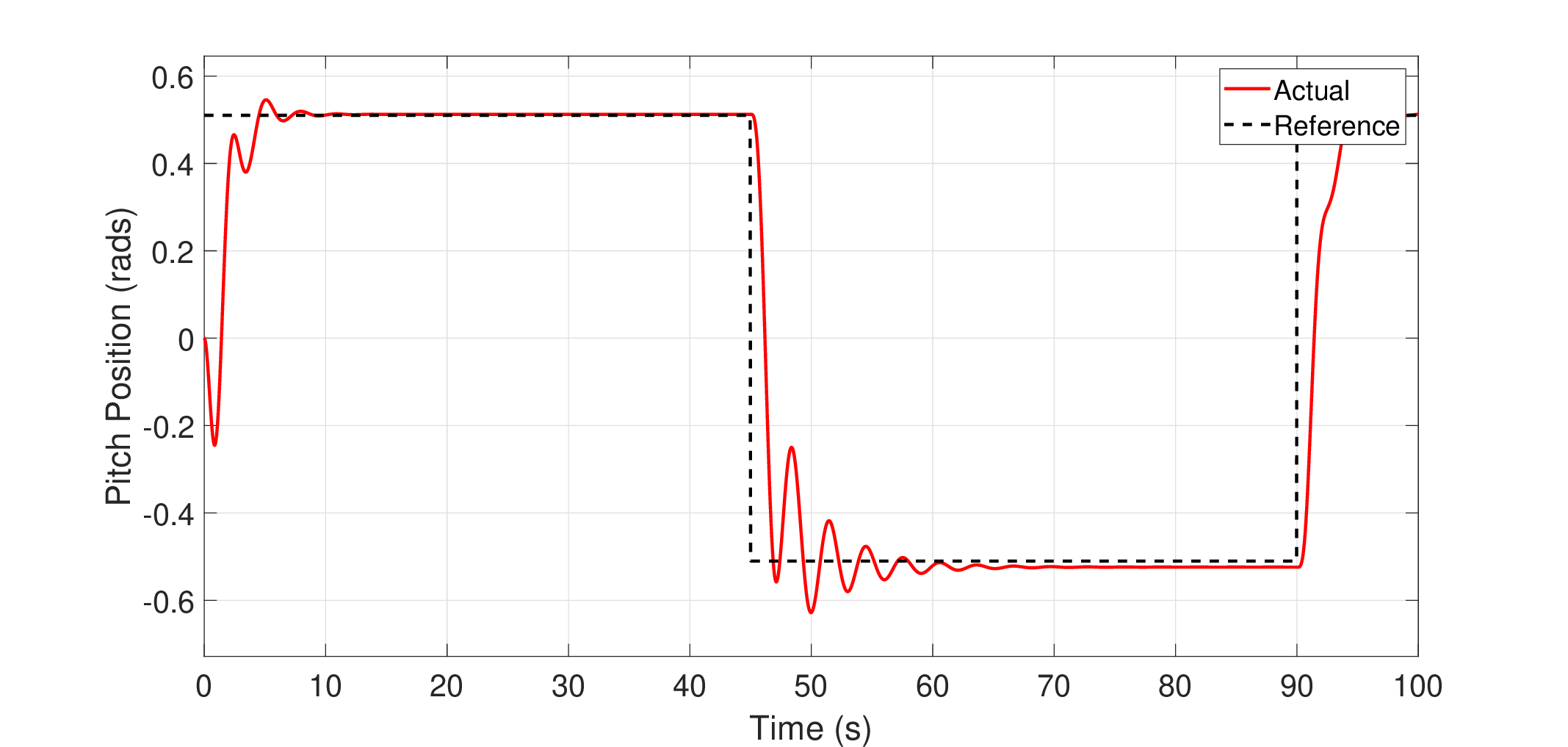}
            \caption{Pitch position (SINDy-TD3)}
            \label{fig:sindy-square-pitch}
        \end{subfigure}
%        \caption*{\textbf{(a)} SINDy-TD3 Square Wave Trajectory Tracking}
    \end{subfigure}

    \vspace{0.2cm}

    % --- Second Row: TD3 Square Wave ---
    \begin{subfigure}{\textwidth}
        \centering
        \begin{subfigure}{0.49\textwidth}
            \centering
            \includegraphics[width=\textwidth]{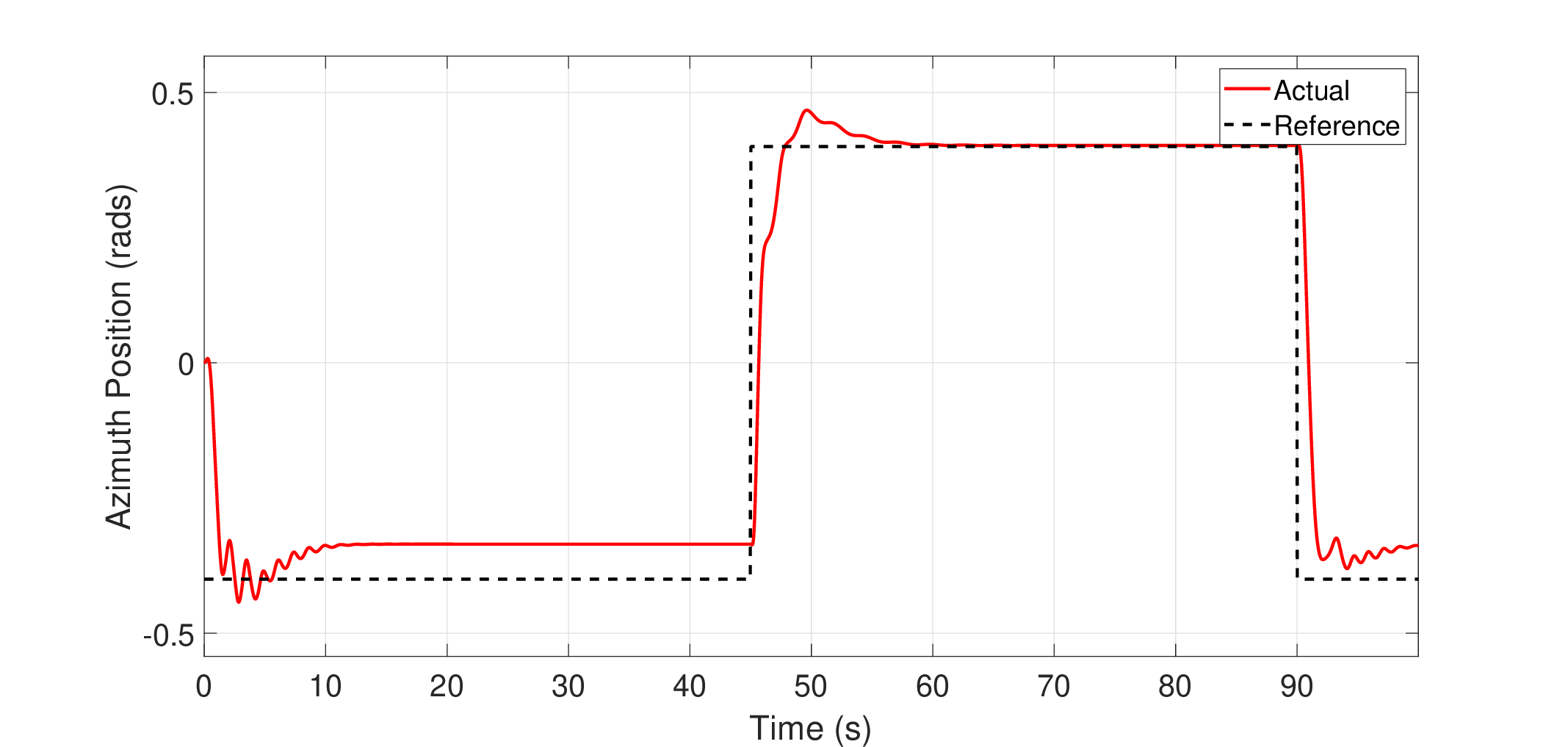}
            \caption{Azimuth position (TD3)}
            \label{fig:td3-square-azimuth}
        \end{subfigure}
        \hfill
        \begin{subfigure}{0.49\textwidth}
            \centering
            \includegraphics[width=\textwidth]{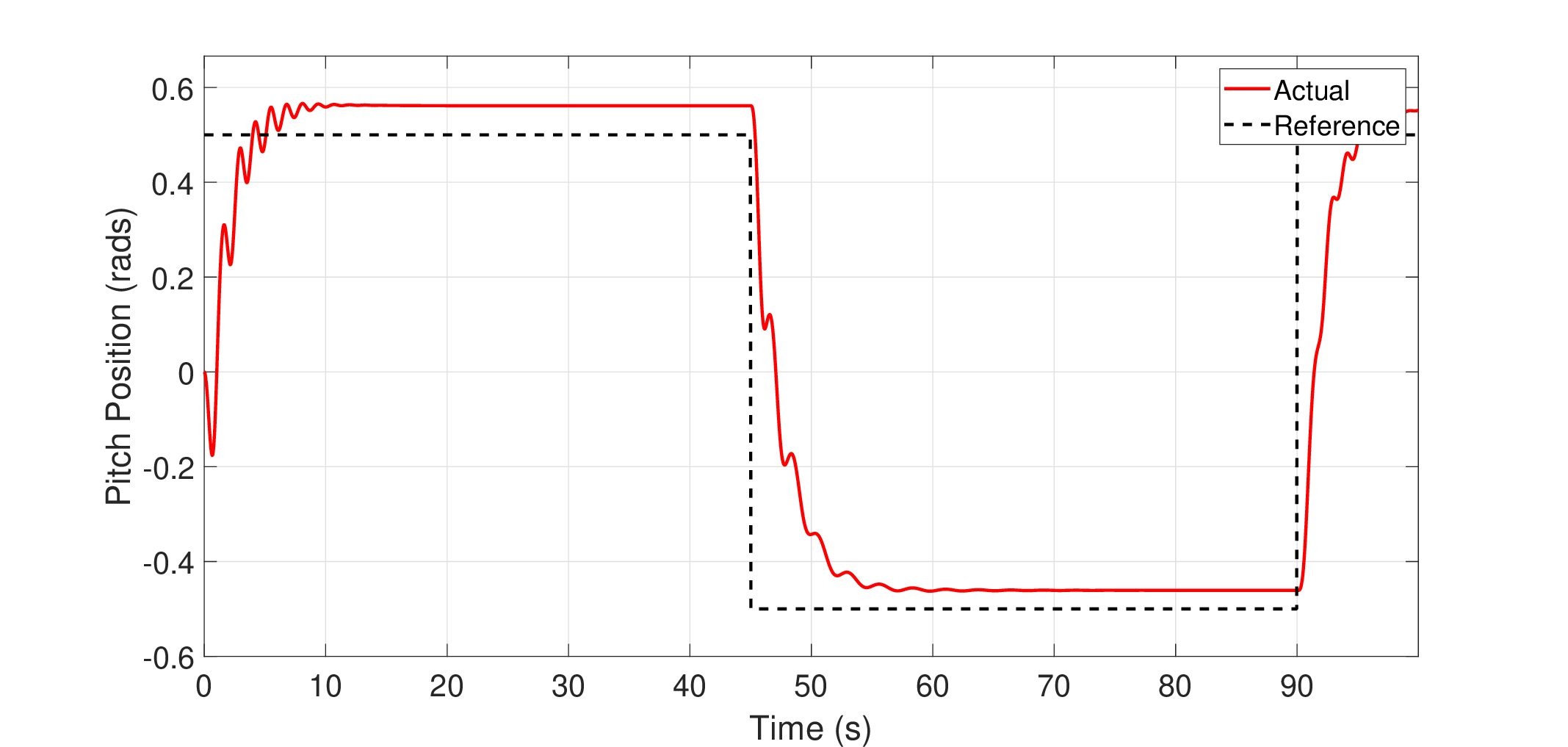}
            \caption{Pitch position (TD3)}
            \label{fig:td3-square-pitch}
        \end{subfigure}
 %       \caption*{\textbf{(b)} TD3 Square Wave Trajectory Tracking}
    \end{subfigure}

    \caption{Comparison between SINDy-TD3 and TD3 in tracking a square wave trajectory for azimuth and pitch angles.}
    \label{fig:comparison-squarewave}
\end{figure}

The results shown in Fig. \ref{fig:comparison-squarewave} (a) and (b) demonstrate the system's ability to successfully track a square wave trajectory for both the azimuth angle and pitch angle positions. The TD3 agent effectively follows the desired trajectory with minimal steady state error, However there are some spikes in the angles due to sudden changes of the reference positions. This highlights DRL robustness and capability in handling trajectory tracking tasks for the bi-rotor system.

The same trajectory was then done however this time with a TD3 agent only without the use of a SINDy environment to validate the accuracy of SINDy-TD3. As shown in Figure \ref{fig:comparison-squarewave} (c) and (d), the TD3 agent was able to accurately track the desired trajectory however, there are more steady state error and less spikes in the response than SINDy-TD3, which again proves the importance of combining SINDy to DRL for data efficiency and less computational power as SINDy provides a simple model yet accurate so that it does not require as much training as TD3 alone where there is no model at all so knowing the dynamics enhances the ability of the TD3 algorithm to provide more accurate results.

\subsection{Discussion}

While model-free RL methods such as TD3 are known for their simplicity and robustness, they typically require large amounts of interaction with the real environment. Conversely, model-based methods, such as those relying on planning, can achieve sample-efficient learning but are sensitive to model inaccuracies. Our approach combines the strengths of both by using a SINDy-learned model to simulate experience, enabling a TD3 agent to train in a data-efficient yet robust manner. This reflects a growing consensus that hybrid approaches, rather than pure model-free or model-based strategies, may offer the best of both paradigms.

A core advantage of the SINDy-TD3 approach lies in its more interpretable structure than model free techniques. The SINDy output is a sparse closed-form differential equation that describes the dynamics of the system \cite{rudy2016datadrivendiscoverypartialdifferential}. Unlike black-box models produced by neural networks which is used in learning simulators, SINDy provide models which are 
physically meaningful that facilitate analytical understanding, debugging, and insight into system behavior. While the control policy itself is still represented by a neural network in the TD3 algorithm, training it with a surrogate environment from SINDy allows greater transparency in how the agent learns and experiences from simulated roll-outs. This makes the whole framework attractive for safety critical or engineering applications where understanding system behavior is significant.

Furthermore, model-free techniques tend to require extensive real-world interactions in order to achieve a good policy, especially in continuous control tasks where small errors in value estimation can only be corrected through repeated exploration \cite{Henderson_Islam_Bachman_Pineau_Precup_Meger_2018}. This leads to heavy demand on real-world systems and long training times, which is not practical or safe. However, SINDy-TD3 benefits from model guided planning in which simulated roll-outs produced by the learned SINDy model replaces a huge amount of real-time physical interactions. As a result, the agent can now perform larger number of training updates in a shorter period, accelerating policy learning and optimization as well as reducing wear, cost, and risk associated with real-world experimentation. This efficiency will be more valuable in settings with limited access to the physical system or where safety constraints severely restrict exploration.

In summary, SINDy-TD3 offers more than just data efficiency, it provides a structured, interpretable, and robust way to learn policies. It brings together the strengths of analytical modeling and deep learning while mitigating many of the weaknesses of model-free RL.

\section{Conclusion}

In this paper, a dyna-style approach for modeling and control of non-linear systems was proposed using SINDy-TD3 to solve the data efficiency problem of model free reinforcement learning. A SINDy data-driven model was acquired from a limited data set to be used to generate simulated rollouts for the TD3 agent reducing its interaction with real world environment.This approach was then tested and validated on a bi-rotor lab setup. The SINDY model was able to capture the Bi-rotor dynamics accurately and hence allowing the TD3 agent to successfully map the bi-rotor states into control inputs in order to let the bi-rotor track a desired trajectory or stabilize in a certain position. Simulations were done on several trajectory scenarios and the agent managed to track all trajectories successfully proving its effectiveness and robustness.

\section{Declarations}

\begin{itemize}
    \item Funding: The authors did not receive support from any organization for the submitted work.
    \item Ethics approval: Not applicable.
    \item Competing Interests: The authors have no conflicts of interest to declare that are relevant to the content of this article.

\end{itemize}

\bibliography{sn-bibliography}% common bib file
%% if required, the content of .bbl file can be included here once bbl is generated
%%\input sn-article.bbl

\end{document}